\documentclass[]{aa}
\usepackage{gensymb}

\usepackage{geometry}
\usepackage{soul}
\usepackage{graphicx}
\usepackage{color}
\usepackage{wrapfig}

\usepackage{txfonts}

\newcommand{\gcc}{\mbox{ g cm$^{-3}$}}

\usepackage{bbold}
\usepackage{caption}
\usepackage{subcaption}

\usepackage{here}
\usepackage{tikz} 
\usepackage[fleqn]{amsmath}
\usepackage{mathrsfs}

\usepackage[colorlinks=True,pdfborder={0 0 0}, linkcolor=blue,citecolor=blue,breaklinks=true]{hyperref}

\usepackage{mwe}

\begin{document}

  \renewcommand{\vec}[1]{\mathbf{#1}}

   \title{Magnetic clumping of charged dust in the dense interstellar medium}

   \author{ V. Vallucci-Goy
          \inst{1} \and
          P. Hennebelle 
          \inst{1} \and
          U. Lebreuilly 
          \inst{1} \and
          G. Verrier.
          \inst{2}
 }

   \institute{Université Paris-Saclay, Université Paris Cité, CEA, CNRS, AIM, 91191, Gif-sur-Yvette, France. \and
   Universit\'{e} Paris Cité, Universit\'{e} Paris-Saclay, CEA, CNRS, AIM, F-91191, Gif-sur-Yvette, France. \\
              \email{valentin.vallucci@gmail.com}
             }

   \date{}

 
  \abstract
   {Dust grains undergo significant growth in star-forming environments, especially in dense regions prone to gravitational collapse. Although dust is generally assumed to represent $1 \%$ of the gas mass, dust density variations are expected on small scales due to differential dynamics with the gas, leading to enhanced coagulation rates in regions of dust enrichment. }
   {We aim to investigate the clumping of charged dust in the turbulent magnetized dense regions of the interstellar medium.}
   {We develop a dusty model that goes beyond the standard non-ideal MHD and use the code {\ttfamily shark} to perform multifluid 1D simulations of a single size charged dust species and neutral gas with large scale driven turbulence and including ion-neutral friction.}
   {Propagating non-linear circularly polarized Alfvén waves, we identify a mechanism similar to the parametric instability that efficiently forms dust clumps even in presence of dissipative processes such as Ohmic dissipation, Hall effect and magnetic drag. Such strong clumping survives and is sustained when driving turbulence, and thus high levels of dust concentration are produced due to compressive magnetic effects in regions of shocks. Dust density enhancements are favored by a high transverse-to-longitudinal magnetic ratio $B_\perp / B_\parallel$ which is controlled by the two following simulation parameters: transverse Mach number $\mathcal{M}_\perp$ and plasma parameter $\beta$. We find that a substantial fraction of dust experiences a density increase of more than a factor of 10 under reasonable conditions (subsonic turbulence, $\beta=0.7$ and dust size $s_\mathrm{d} \geq 1 \ \mathrm{\mu m}$), thus promoting dust growth.}
   {Our novel dusty non-ideal MHD model shows that dust grains (main charge carriers) are subject to small-scale compressive magnetic effects driven by a parametric instability - like mechanism in regions of shocks, and consequently experience high density enhancements in turbulent environments that go beyond those permitted by pure hydrodynamical processes, making in-situ formation of large grains ($s_\mathrm{d} \sim 100 \ \mathrm{\mu m}$) in protostellar envelopes a plausible scenario.  
 
   }


    \authorrunning{Vallucci-Goy et al.}
    \titlerunning{Magnetic clumping of charged dust in the dense interstellar medium}

    \maketitle

%
\section{Introduction}

Interstellar dust is of great importance in many astrophysical environments in regard to the thermodynamics and chemistry of its surrounding medium.
In particular, dust grains are essential ingredients for star and planet formation. The wide spectrum of dust grain sizes affects the optical properties of the medium and, consequently, the heating and cooling processes at work \citep{McKee2007}. In addition, dust is a very useful observational probe in the interstellar medium (ISM) and more generally in the context of star and planet formation \citep{Draine2004}. Dust grains affect the ionization degree of the medium and can become the main charge carriers during protostellar collapse \citep{Zhao2016,Tuskamoto2022} as they allow for electron and ion recombination and electron capture on their surface \citep{Ivlev2015,Marchand2016}. As such, they control the degree of coupling between the gas and the magnetic field \citep{Nakano2002}. All of the above processes are sensitively dependent on local dust concentration levels and on dust size which is modelled as a power-law Mathis-Rumple-Nordsiek (MRN) distribution in the diffuse ISM \citep{Mathis1977}. However, the dust size distribution is expected to evolve as dust grains undergo coagulation and fragmentation processes \citep{Dominik1997,Ormel2009}. Following such an evolution of the dust distribution is very challenging, but essential. Dust growth is also inherently related to the formation of larger solid bodies such as planetesimal and later on planets \citep[presumably in less than 1 Myr, see][]{Testi2014}.

Dust growth is very sensitive to the amount of dusty material available, i.e. to dust density. Although it is widely accepted that dust represents only $1 \%$ of the gas mass in the ISM and in protostellar dense cores, this average value holds on large scale but could vary on smaller scales due to various mechanisms such as gas-dust differential dynamics \citep{Lebreuilly2020}, turbulence and (magneto)-hydrodynamical (MHD) instabilities \citep[see ][respectively for Kelvin-Helmholtz instability and resonant drag instabilities (RDI)]{Hendrix2014,Squire2018}. A number of works have recently studied such concentration of solid particles, unveiling up to orders of magnitude in dust-to-gas ratio fluctuations resulting from aerodynamical drag. Hydrodynamical dust clumping by supersonic turbulence has been studied numerically in an environment reproducing molecular cloud physical conditions \citep{Hopkins2016,Tricco2017,Commerçon2023}. In the latter, they compare two numerical implementations, namely a monofluid Eulerian approach with terminal velocity approximation for the dust grains, and a bifluid approach treating the dust grains as Lagrangian superparticles. They show that the terminal velocity approximation is well suited for dense regions but dust decoupling and enrichment is not well described for grains of Stokes number above unity (in regions of low density). In the same Stokes number regime, the bifluid approach capacity to describe dust enrichment is limited by the grid resolution used for the gas. For very well coupled dust grains ($\mathrm{St} << 1$), the Lagrangian treatment of the dust leads to artificial trapping in the densest regions \citep{Price&Federrath2010,Cadiou19}. Large grains tend to clump more efficiently as they are less coupled to gas molecules \citep{Mattsson2019}. With both an analytical and numerical approach, \cite{Hopkins2018} expanded their work on the resonant drag instability (RDI) by including the role of magnetic fields and charged dust. As dust and gas drift relative to one another (via a different force balance including, for example, gravity for both and pressure gradients for the gas only), instabilities develop for all mode wavelengths, leading to strong fluctuations in the dynamics of both fluids. In later phases of the star formation cycle, such as protoplanetary disks, a number of mechanisms and instabilities have been suggested and studied to significantly concentrate the dust locally \citep{Xu&Bai2022,Lehmann2023,Birnstiel2024}, including the so-called Streaming Instability \citep[][and references therein]{Youdin2005,Johansen2007,Drazkowska2017,Krapp2019,LiYoudin2021,Carrera2025a} and the Dust Settling Instability \citep[part of the RDI family, see][]{Squire2018,Krapp2020a}.

Including magnetic fields, \cite{Lee&Hopkins2017,Moseley2023,moseley2025} have investigated dust concentration in molecular clouds considering charged dust as Lagragian particles feeling Lorentz forces. In such diffuse environments, most of the charges are mobile ions and electrons and the authors assumed them to be moving as one with gas molecules, therefore solving the induction equation in the ideal regime to compute the magnetic field evolution. In spite of considering charged dust grains, magnetic field evolution is dictated by gas motions, i.e. field lines are attached to the gas fluid in the induction equation. They find initially very large dust-to-gas ratio fluctuations being curtailed to some extent for small particles by Lorentz forces, which offer an extra coupling between gas and dust, while large grains are less affected due to their lower magnetization. \newline
However, to our knowledge, studies of dust clumping with charged dust and beyond-standard MHD models in denser environments of the ISM  such as protostellar envelopes and dense cores have never been undertaken. At such densities ($> 10^{4} \ \mathrm{cm^{-3}}$), small dust grains are to be considered as main charge carriers \citep[][]{Nishi1991,Nakano2002,Zhao2016} as they collect electrons and recombine positive and negative charges on their surface. More sophisticated MHD models and chemical networks have to be used to model correctly dust and gas dynamics. In this work, we perform 1D simulations of a two-fluid (magnetized dust and neutral gas) mixture in a turbulent box, using lower computational cost due to our lack of multidimensionality to focus on microphysical processes and clumping on very small scale.
In Sect. \ref{methods}, we present the methods and MHD equations solved. Numerical setups are described in Sect. \ref{Setups}. In Sect. \ref{Results}, we first consider a single non-linear torsional Alfvén wave, emphasizing the role of a parametric-like instability to produce compressible modes (note that we let numerical noise serve as perturbations). Then, we test the robustness of the presented clumping mechanism in a more realistic environment including 1D turbulence driven on large scales. We discuss those results in Sect. \ref{discussion}. Finally, we summarize our findings in Sect. \ref{Section conclusion} and provide additional material in the Appendix.

\section{Methods: multifluid implementation}
\label{methods}

We use the code {\ttfamily shark} \citep{Lebreuilly2023} to run our simulations. We consider a 1D box with a uniform grid and periodic boundary conditions which extends along the x-axis. We make a full treatment of the gas and dust dynamics which interact through drag terms. The gas evolves isothermally and the dust fluid is considered pressureless. The box is threaded by a uniform magnetic field. We allow transverse (y and z-wise) components to exist.

\subsection{Dusty non-ideal MHD}
 \label{Dusty non-ideal MHD}

 Considering neutral gas and a single charged dust fluid, the magnetohydrodynamical equations are:

 \begin{align}
 \frac{\partial \rho}{\partial t}  + \nabla \cdot \left[ \rho \vec{v} \right]&=&0,  \nonumber \\   \frac{\partial   \rho_\mathrm{d} }{\partial t} +\nabla \cdot \left[ \rho_\mathrm{d} \vec{v_\mathrm{d}} \right] &=&0, \nonumber \\
  \frac{\partial \rho \vec{v}}{\partial t}  + \nabla \cdot \left[ \rho \vec{v} \vec{v} + P \mathbb{I} \right]&=& 
  \frac{\rho_\mathrm{d}}{t_{\rm{s},d}}  (\vec{v}_\mathrm{d}-\vec{v}) + \vec{F_\mathrm{i \rightarrow N}}, \nonumber\\
  \frac{\partial \rho_\mathrm{d} \vec{v}_\mathrm{d}}{\partial t}  + \nabla \cdot \left[ \rho_\mathrm{d} \vec{v}_\mathrm{d} \vec{v}_\mathrm{d} \right]&=&   - \frac{\rho_\mathrm{d}}{t_{\rm{s},d}} (\vec{v}_\mathrm{d}-\vec{v}) + \vec{F_\mathrm{Lor,d}},
  \label{eq:hydro}
\end{align}
where the first term on the right hand side of the momentum equations is the hydrodynamical drag between gas and dust. $\rho$ and $\rho_\mathrm{d}$ refer respectively to the gas and dust density. $\vec{F}_\mathrm{Lor,d} = n_\mathrm{d} Z_\mathrm{d} e \left(\vec{E} + \frac{\vec{v_\mathrm{d}}}{c} \times \vec{B} \right)$ is the Lorentz force felt by the dust fluid. $n_\mathrm{d}$ and $Z_\mathrm{d} < 0$ are the dust numerical density and the number of electrical charges carried by dust grains. $e$ is the fundamental electrical charge, $c$ the speed of light and $\vec{E}$ the electric field. We also include ions, whose inertia is neglected but we ignore the presence of electrons for the sake of simplicity. We consider the friction exerted by ions on neutrals to be $\vec{F_\mathrm{i \rightarrow N}} = \rho_\mathrm{i} \rho \gamma_\mathrm{i,g} \left(\vec{v_\mathrm{i}} - \vec{v} \right) $. The ions momentum equation is:

\begin{align}
\label{ion mom eq}
  e n_\mathrm{i} \left(\vec{E} + \frac{\vec{v_\mathrm{i}}}{c} \times \vec{B} \right) - \vec{F_\mathrm{i \rightarrow N}} &=& 0, 
\end{align}
where $n_\mathrm{i}$ is the ion numerical density.

We use the momentum equations of the ions along with electroneutrality condition to infer a generalized Ohm's law, which gives us the electric field (the complete derivation is provided in the Appendix, Sect. \ref{Generalized Ohm's law}). We then inject the electric field $\vec{E}$ in the induction equation to get:
\begin{align}
\label{induction equation}
        \frac{\partial \vec{B}}{\partial t} - \nabla \times \left[\vec{v}_\mathrm{d} \times \vec{B} - \frac{\|\vec{B}\|}{\Gamma_\mathrm{i}} \left(\vec{v_\mathrm{d}} - \vec{v}\right) \right]  = \nabla \times \left[\frac{c}{e n_\mathrm{i} 4 \pi} \left(\nabla \times \vec{B} \right) \times \vec{B} \right]  \nonumber \\ 
        - \nabla \times \left[\frac{c \|\vec{B}\|}{\Gamma_\mathrm{i} e n_\mathrm{i} 4 \pi} \left(\nabla \times \vec{B} \right) \right],
\end{align}
 where the first term on the right hand side takes the form of a dispersive Hall term and the second of a dissipative Ohm term. The former is treated as a conservative term while the latter is treated as a diffusion equation. More information along with a test of our solver for the treatment of diffusion equations are provided in Sect. \ref{Diffusion test section}. Note also that the Hall effect makes almost no difference in our simulations. We define the ion Hall factor $ \Gamma_\mathrm{i} = \omega_i t_\mathrm{s,i}$ to be the product of the ion stopping time with the gyromagnetic frequency $\omega_i = \frac{e \|\vec{B}\|}{cm_i}$. The ion stopping time $t_\mathrm{s,i} = \left(\rho \gamma_\mathrm{i,g} \right)^{-1}$ is given in \cite{Pinto2008} and \cite{Marchand2016} for $\mathrm{HCO^+}$ ions.
 As mentioned earlier, our box is threaded by a uniform background magnetic field $B_x = constant$. Within this 1D configuration, the divergence-free condition $\nabla \cdot \vec{B} = \partial_x B_x = 0$ is naturally complied and no constrained transport scheme is needed.
 
We also inject the electric field $\vec{E}$ in the dust Lorentz force and obtain after a few calculation steps (Sect. \ref{Generalized Ohm's law}):
\begin{equation}
\label{Dust Lor Force}
\vec{F_\mathrm{Lor,d}} = \frac{\vec{J_\mathrm{tot}}}{c} \times \vec{B} - \frac{\|\vec{B}\|}{\Gamma_\mathrm{i} c} \vec{J_\mathrm{tot}} - \frac{e n_\mathrm{d} Z_\mathrm{d} \|\vec{B}\|}{\Gamma_\mathrm{i} c}(\vec{v}-\vec{v}_\mathrm{d}),
\end{equation}
 where $\vec{J_\mathrm{tot}} = \vec{J_\mathrm{d}} + \vec{J_\mathrm{i}}$ is the total electric current (contributions from dust and ions) related to the magnetic field as per Maxwell-Ampère equation:

 \begin{equation}
 \label{Maxwell-Ampère}
\vec{J_\mathrm{tot}} = \frac{c}{4 \pi} \nabla \times \vec{B}.
\end{equation}

The first term of Eq. (\ref{Dust Lor Force}) has to be rewritten in conservative form and moved to the left-hand part of the dust momentum equation to ensure stability of the dust Riemann solver. The third term behaves as an additional friction term between gas and dust coming from the presence of ions. This yields:

\begin{align}
\label{final dust mom eq}
    \frac{\partial \rho_\mathrm{d} \vec{v_d}}{\partial t}  + \nabla \cdot &\left[ \rho_\mathrm{d} \vec{v}_\mathrm{d} \vec{v}_\mathrm{d} + \frac{B^2}{8 \pi} - \frac{\vec{B}\vec{B}}{4 \pi} \right] = \\ \nonumber
&\quad -\frac{\|\vec{B}\|}{\Gamma_\mathrm{i} c} \vec{J_\mathrm{tot}} - \frac{e n_\mathrm{d} Z_\mathrm{d} \|\vec{B}\|}{\Gamma_\mathrm{i} c}(\vec{v}-\vec{v}_\mathrm{d}) + \frac{\rho_\mathrm{d}}{t_{\rm{s},d}} (\vec{v}-\vec{v}_\mathrm{d}).
\end{align}
Similarly for the gas, we use Eq. (\ref{ion mom eq}) to substitute $\vec{F_\mathrm{i \rightarrow N}}$ with $e n_\mathrm{i} \left(\vec{E} + \frac{\vec{v_\mathrm{i}}}{c} \times \vec{B} \right)$ and get:
\begin{align}
    \frac{\partial \rho \vec{v}}{\partial t}  + \nabla \cdot &\left[ \rho \vec{v} \vec{v} + P \mathbb{I} \right] = \\ \nonumber
&\quad \frac{\|\vec{B}\|}{\Gamma_\mathrm{i} c} \vec{J_\mathrm{tot}} - \frac{e n_\mathrm{d} Z_\mathrm{d} \|\vec{B}\|}{\Gamma_\mathrm{i} c}(\vec{v}_\mathrm{d}-\vec{v}) + \frac{\rho_\mathrm{d}}{t_{\rm{s},d}} (\vec{v}_\mathrm{d}-\vec{v}).
\end{align}

Both drag terms between gas and dust are treated implicitly using the solver presented in \cite{Krapp2020b}. We assume the mean free path of gas molecules to be large compared with the dust grain size. This is known as the Epstein regime \citep{Epstein1924} for which the stopping time of the grain is expressed as:
\begin{equation}
\label{ts}
    t_\mathrm{s,d} = \sqrt{\frac{\pi}{8}} \frac{\rho_\mathrm{gr}s_\mathrm{d}}{\rho c_s},
\end{equation}
where $\rho_\mathrm{gr} = 2.3 \ \gcc $ is the internal density of a (spherical) dust grain and $s_\mathrm{d}$ its size (diameter). $c_\mathrm{s}$ is the isothermal sound speed. $t_\mathrm{s,d}$ is the characteristic time needed for a dust grain to adjust its velocity to a change in gas velocity. We define the dimensionless Stokes number $\mathrm{St}$ as:
\begin{equation}
  \mathrm{St} =   t_\mathrm{s,d} / t_\mathrm{dyn} =t_\mathrm{s,d} c_s / L.
\end{equation}
The dynamical timescale of interest $t_\mathrm{dyn}$ is taken to be the time required for a sound wave to cross our numerical box of length $L$. $\mathrm{St} = \mathrm{St_0} \rho_0/\rho$ where $\mathrm{St_0}$ is the initial Stokes number (associated with the background state of initial uniform gas density $\rho_0$ and scale $L$).

\subsection{Dusty ideal MHD: perfect coupling between dust and magnetic field}
\label{dusty ideal MHD}

If we neglect ion-neutral friction and have the ion Hall factor $\Gamma_\mathrm{ion} \longrightarrow \infty$, as well as drop the Hall term in the induction equation, the latter takes the classic form encountered in ideal MHD, however with the dust velocity in place of the gas velocity:

\begin{equation}
\label{induction equation ideal}
        \frac{\partial \vec{B}}{\partial t} - \nabla \times \left[\vec{v_d} \times \vec{B} \right]  = 0.
\end{equation}
The dust fluid is then considered as the sole charge carrier and is frozen to the magnetic field lines. The gas and dust momentum equations become:

 \begin{align}
  \frac{\partial \rho \vec{v}}{\partial t}  + \nabla \cdot \left[ \rho \vec{v} \vec{v} + P \mathbb{I} \right]&=& 
  \frac{\rho_\mathrm{d}}{t_{\rm{s},d}}  (\vec{v}_\mathrm{d}-\vec{v}), \nonumber\\
  \frac{\partial \rho_\mathrm{d} \vec{v_d}}{\partial t}  + \nabla \cdot \left[ \rho_\mathrm{d} \vec{v}_\mathrm{d} \vec{v}_\mathrm{d} + \frac{B^2}{8 \pi} - \frac{\vec{B}\vec{B}}{4 \pi} \right]&=&   - \frac{\rho_\mathrm{d}}{t_{\rm{s},d}} (\vec{v}_\mathrm{d}-\vec{v}),
  \label{eq:hydro ideal}
\end{align}
This approach is used to construct the dispersion relation depicted in Sect. \ref{Alfven propagation section}.

\subsection{Ionization}
\label{chemical network}
In order to compute the charge abundances and stay consistent with the assumptions made within our MHD model, we need to neglect the presence of electrons in our chemical network (in the electroneutrality condition). To avoid further complicating the physics, we use a very simple approach where the abundance of ions ($\mathrm{HCO^{+}}$) is given by a balance between cosmic-ray ionization (assuming an ionization rate of $\zeta = 1 \times 10^{-17} \ \mathrm{s^{-1}}$) and two-body recombination of charged particles \citep{Elmegreen1987,Shu1987}:

\begin{equation}
\label{Shu}
    x_i = \frac{n_\mathrm{i}}{n_\mathrm{H}} = 10^{-7} \left(\frac{n_\mathrm{H}}{10^{3} \ \mathrm{cm^{-3}}}\right)^{-1/2},
\end{equation}
and the electroneutrality condition is:

\begin{equation}
\label{electroneutrality}
    n_\mathrm{i} + n_\mathrm{d}Z_\mathrm{d} = 0.
\end{equation}

\section{Numerical setup}
\label{Setups}

\subsection{Initial conditions}

Our setup is intended for the exploration of dust clumping in a magnetized turbulent environment, such as protostellar dense cores and envelopes. Our 1D box extends only in the x direction, but we allow transverse components (y and z-wise) of magnetic field and velocities to exist as well. The simulations are initiated with uniform gas density $\rho_0$ and dust density such that $\rho_\mathrm{d,0} = 0.01 \rho_0$ (i.e., a dust-to-gas ratio $\epsilon = 0.01$), a uniform x-wise background magnetic field inferred from the plasma parameter $\beta = c_\mathrm{s}^2/c_\mathrm{a}^2$ ($c_\mathrm{a}$ being the Alfvén velocity), which gives:

\begin{equation}
\label{plasma beta}
B_x = B_\parallel = c_\mathrm{s} \sqrt{4 \pi \rho_0/\beta},
\end{equation}
and a constant isothermal sound-speed $c_\mathrm{s}  = \sqrt{\frac{ k_{\mathrm{B}}T}{\mu m_{\rm{H}}}}$ with a temperature $T = 10 \ \mathrm{K}$ (because gas is rapidly cooling). $\mu = 2.3$ is the mean molecular weight of the gas molecules in terms of units of hydrogen mass $m_\mathrm{H}$. Introducing the gas numerical density $n_\mathrm{H} = \rho / \mu m_\mathrm{H}$, the box length $L$ is taken equal to the Jeans length which is defined as the distance crossed by a sound wave in a free-fall time:

\begin{equation}
    L = c_st_\mathrm{ff} = \frac{c_s}{\sqrt{\mathcal{G} \mu m_\mathrm{H}n_\mathrm{H}}}. 
\end{equation}

\subsection{Driven turbulence}

We drive longitudinal (x-wise) and transverse (y and z-wise) turbulent motions in the gas by adding an acceleration kick at each timestep, composed of 10 sinusoidal modes of wavenumber lying in the range $L/3 \leq k_\mathrm{mode,i} \leq L/2$ (i.e., the turbulence is injected on a scale which is a fraction of the Jeans length):

\begin{equation}    
\label{turb kick}
    F_\mathrm{kick,xyz} = \rho \sum_\mathrm{i=1}^{10} A_\mathrm{xyz,i} \sin{\left(k_\mathrm{mode,i} x + \phi_\mathrm{xyz,i}\right)} 
\end{equation}
The amplitude $A_{i}$ of each mode is generated randomly from a Gaussian distribution and we allow the phase to randomly (from uniform distribution) drift over a turnover time taken to be equal to the sound crossing time, i.e. $t_\mathrm{turnover} = t_\mathrm{cross} = L/c_\mathrm{s}$.
We measure the Mach number through the volume-averaged root mean squared (RMS) velocity $v_\mathrm{RMS}^2 = \int_0^L \left(\vec{v}-\overline{\vec{v}} \right)^2dx/\int_0^L dx$, (with $\overline{\vec{v}} = \int_0^L \vec{v} dx / \int_0^L dx$) and calibrate the modes amplitude with a correcting factor to approach equipartition and to vary the  Mach number when needed. We do this for each component, which leads to the total Mach number $\mathcal{M} = v_\mathrm{RMS}/c_\mathrm{s}$ where $v_\mathrm{RMS}^2 = \int_0^L \left({v_x}-\overline{{v_x}} \right)^2dx/\int_0^L dx + \int_0^L \left({v_y}-\overline{{v_y}} \right)^2dx/\int_0^L dx + \int_0^L \left({v_z}-\overline{{v_z}} \right)^2dx/\int_0^L dx = v_\mathrm{x,RMS}^2 + v_\mathrm{y,RMS}^2 + v_\mathrm{z,RMS}^2$. It follows that $\mathcal{M}^2 = \left(v_\mathrm{x,RMS}^2+v_\mathrm{y,RMS}^2+v_\mathrm{z,RMS}^2\right)/c_\mathrm{s} = \mathcal{M_\parallel}^2 + \mathcal{M_\perp}^2$ where we denote $\mathcal{M_\parallel} = \mathcal{M}_x$ the Mach number in the longitudinal direction (x-wise) and $\mathcal{M_\perp} = \sqrt{\mathcal{M}_y^2 + \mathcal{M}_z^2}$ the transverse one, perpendicular to the background uniform magnetic field $B_x$.  
Both the dust and gas are initially at rest.

\section{Results}
\label{Results}

In this section, we investigate dust and gas clumping.
We first provide a known and documented framework without turbulence in order to give a simple description of the clumping mechanism observed in our simulations. The remainder of this section is then dedicated to the influence of previously mentioned parameters in simulations with turbulence and to the question whether dust clumping endures in such a turbulent environment.

\subsection{Dust clumping due to a parametric-like instability}
\label{section parametric instability}

\begin{figure*}[hbt]
    \sidecaption
    \includegraphics[width=12cm]{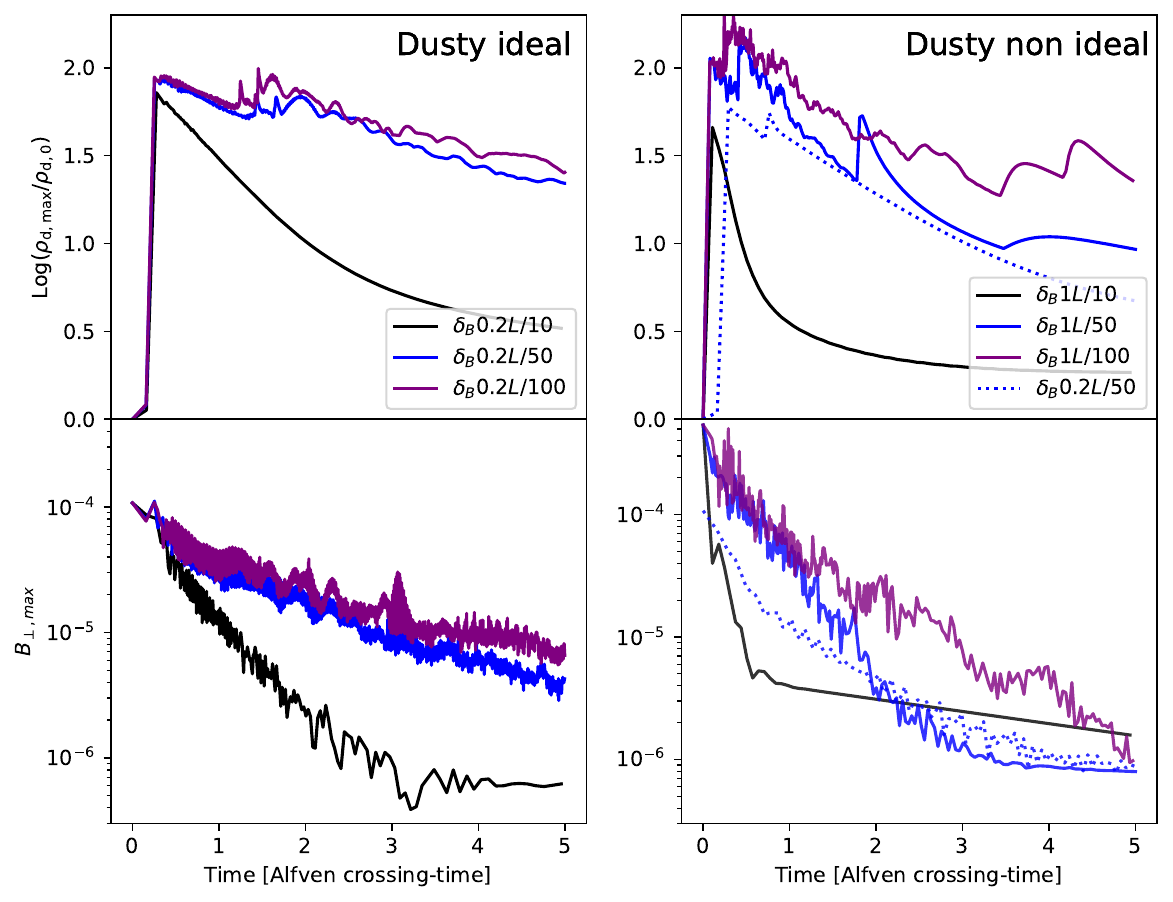}
    \caption{Maximum dust density fluctuations and maximum transverse magnetic field as a function of time upon propagation of a circularly polarized Alfvén wave, for different values of wavelength (as a fraction of the box length $L$) and wave amplitudes $\delta_B$ within the dusty ideal MHD regime and dusty non-ideal MHD regime (see Sect. \ref{dusty ideal MHD} and Sect. \ref{Dusty non-ideal MHD}). The early sharp rise in dust density is due to a mechanism similar to the parametric instability \citep{DelZana2001}. Note that growth rates measured here compare well with analytical predictions of the standard parametric instability (see Sect. \ref{Appendix: growth rate parametric}). Afterwards, the dust density decreases as a consequence of gas dust friction. The gas density is not displayed because subject to negligible variations. Parameters: $n_\mathrm{H} = 10^6 \ \mathrm{cm^{-3}}$, $s_\mathrm{d} = 10 \ \mathrm{\mu m} \ (\mathrm{St} = 0.01)$ and $\beta = 0.1$.}
    
\label{param instability ideal vs non-ideal}
\end{figure*}

\begin{figure}
\centering
\includegraphics[width=0.5\textwidth]{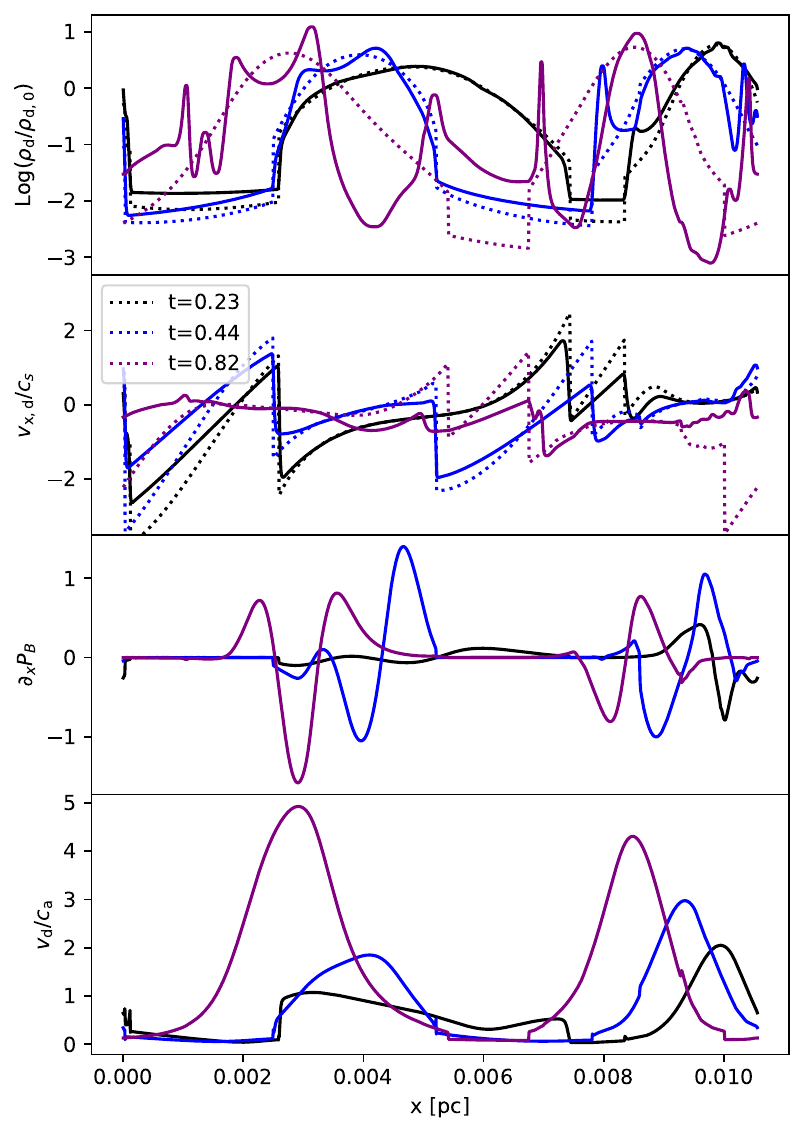}

\caption{Different fields at times $t=0.23$, $t=0.44$ and $t=0.82$ in a simulation with driven turbulence. First panel: dust (solid lines) and gas (dotted lines) density fluctuations. Second panel: x-wise dust (solid) and gas (dotted) Mach number. Third panel: magnetic pressure gradient. Fourth panel: total (x,y,z-wise) dust Alfvénic Mach number. Parameters: $n_\mathrm{H} = 10^6 \ \mathrm{cm^{-3}}$, $s_\mathrm{d} = 10 \ \mathrm{\mu m} \ (\mathrm{St} = 0.01)$ and $\beta = 1$.}

\label{Fields at different times}
\end{figure}

\begin{figure}
\centering
\includegraphics[width=0.4\textwidth]{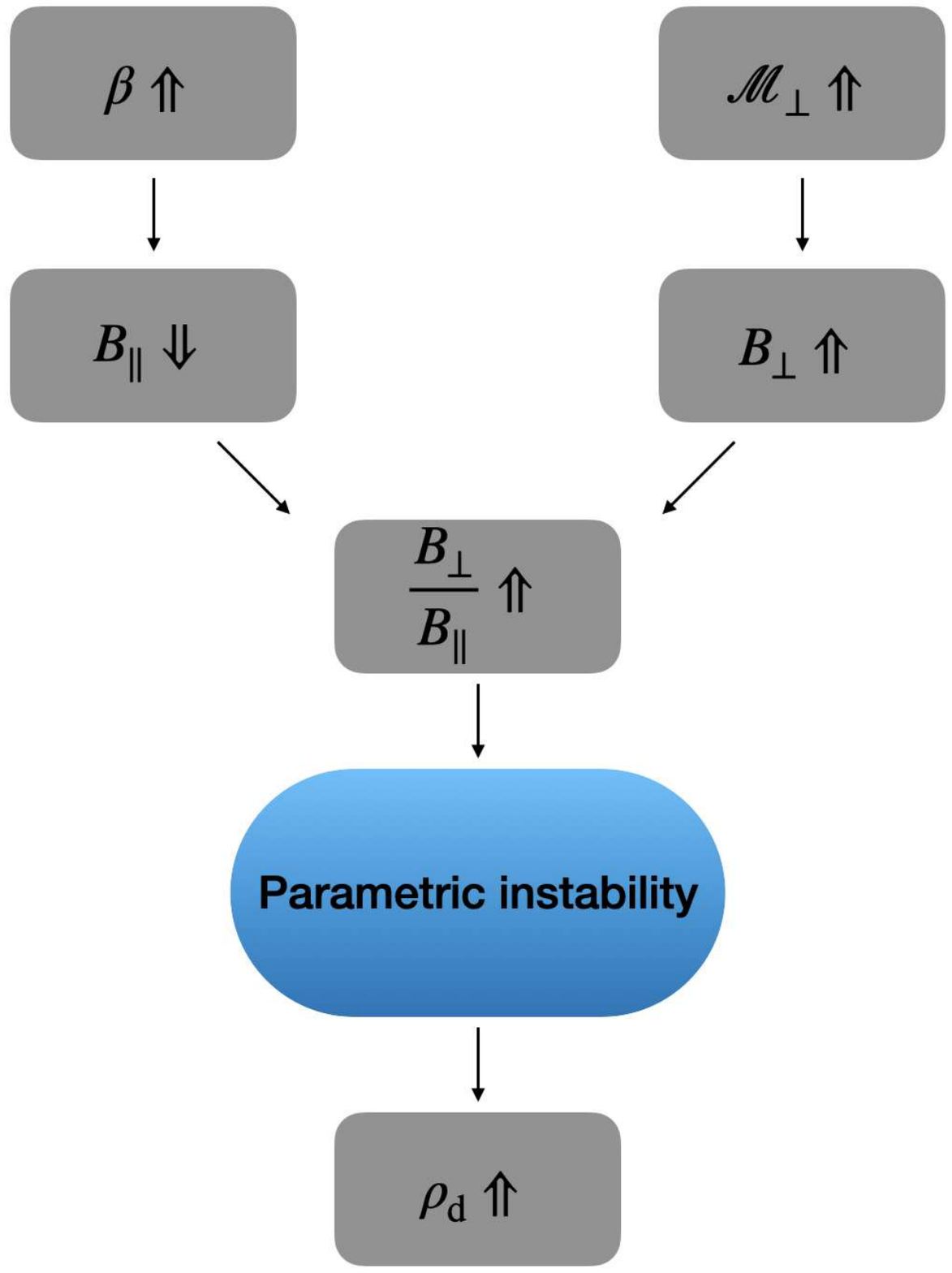}

\caption{Schematic illustration of the parameters controlling the transverse-to-longitudinal magnetic field ratio $B_\perp / B_\parallel$ and thus the intensity of dust clumping via the parametric instability in the simulations with turbulence.}

\label{param instability cartoon}
\end{figure}

For the sake of simplicity, we initiate a non-linear circularly polarized Alfvén wave of the form:

\begin{equation}
B_y = \delta_B B_x \sin{\left(\frac{2 \pi }{\lambda} x\right)} \\  
B_z = \delta_B B_x \cos{\left(\frac{2 \pi }{\lambda} x\right)},
\end{equation}
where the amplitude of the wave is taken to be a fraction $\delta_B$ of the background magnetic field $B_x$. We find the so-called parametric instability \citep[][]{DelZana2001,Hennebelle&Passot2006} to likely be the underlying mechanism responsible for the magnetically induced clumping of the dust. Figure \ref{param instability ideal vs non-ideal} shows dust maximum density fluctuations as a function of time upon propagation of the Alfvén wave, for different values of wavelength (taken to be a fraction of the box length, either $L/10$, $L/50$ or $L/100$) within both the dusty ideal MHD regime (presented in Sect. \ref{dusty ideal MHD}) and the dusty non-ideal MHD regime (presented in Sect. \ref{Dusty non-ideal MHD}). Note that perturbations of the background state come from numerical noise.

The stability of Alfvén waves was first studied long ago by \cite{Galeev1963} and then by many authors. Later, the parametric instability was investigated numerically in the context of a single fluid in the ideal MHD regime with 1D to 3D simulations by \cite{DelZana2001} who showed that a circularly polarized "mother" Alfvén wave (which is an incompressible mode since its magnetic pressure is constant) is unstable and couples to density and compressible fluctuations in the plasma provided that its amplitude is large enough (non-linear regime). Under such circumstances, the wave decays and gives rise to a whole spectrum of "daughter" waves, namely Alfvén modes and compressible modes (prone to non-linear steepening and shock dissipation) propagating in both directions. Such compressible modes feed on the energy of the mother wave and lead to remarkable density fluctuations, sensitive to the frequency and amplitude of the mother wave. \newline
In the context of our two-fluid work, triggering the instability is much more challenging since the magnetized species is the dust, which is altogether 100 times less massive than neutral gas ($\epsilon = 1 \%$). Due to hydrodynamical drag, dust gyro-motions and propagation of Alfvén waves are altered, and are even completely curtailed for certain frequencies (this very dissipative regime is described in the Appendix Sect. \ref{Alfven propagation section} within the dusty ideal MHD setup). When considering dust inertia, propagation is recovered at sufficiently high frequencies \citep{Hennebelle2023}. Looking at Fig. \ref{param instability ideal vs non-ideal} in the dusty ideal regime, we see an early sharp increase in dust density followed by a smoother decrease due to hydrodynamical drag (gas is almost not affected for it is not magnetized). In addition, there is a clear dependency of dust density fluctuations on wavelength. Dust size was chosen to give a Stokes number of $\mathrm{St} = 0.01$, meaning that dust should strongly decouple hydrodynamically from the gas on scales smaller than a hundredth of the box length ($\lambda \leq L/100$). Indeed, while dust density fluctuations drop below $10^{0.75}=5.62$ over a few Alfvén crossing times for $\lambda \geq L/10$, they remain much higher for shorter wavelengths, showing that in this case dust grains are free to concentrate locally. We point out that in the dusty ideal regime, strong dust density enhancements are observed ($\rho_\mathrm{d,max}/ \rho_\mathrm{d,0} \geq 10^{1.5} = 31.6$) in spite of a modest wave amplitude of $\delta_B = 0.2$. Once the decoupling scale has been reached, further reducing the wavelength only leads to a modest increase in maximum dust density as can be seen when comparing $\lambda = L/50$ to $\lambda = L/100$. In addition, the decrease in maximum transverse magnetic field indicates that the Alfvén wave is dissipated due to dust-gas collisions, with a more severe dissipation for large wavelengths.

In addition to hydromagnetic drag, dissipative non-ideal MHD effects (see Sect. \ref{Dusty non-ideal MHD}) further complicate the development of such an instability. Considering now the non-ideal MHD regime in Fig. \ref{param instability ideal vs non-ideal}, it is clear again that reducing the wavelength leads to a higher saturation density as hydrodynamical friction is reduced on smaller scales (when wave frequency starts exceeding the collision frequency). However, magnetic drag will persist even on small scales since the ion stopping time $t_\mathrm{s,i}$ is very low compared to that of dust (see Sect. \ref{section non-ideal effects and chemical network}), leading to overall dust density enhancements lower than in the dusty ideal MHD regime (comparing both models with $\delta_B = 0.2$). In addition, increasing the wave amplitude $\delta_B$ does yield a higher level of dust concentration, but essentially only at early times ($t < 1$) before magnetic drag disrupts dust clumps. A key ingredient to the development of the parametric instability is the transverse-to-longitudinal magnetic field ratio $B_\perp / B_\parallel$ which is here directly controlled by $\delta_B$ (in simulations with turbulence, the parameters controlling this ratio are depicted in Fig. \ref{param instability cartoon}). Similarly, we see that again the transverse magnetic field intensity drops more steeply when considering larger wavelengths. Within the dusty non-ideal regime, the transverse magnetic field approaches a final value that is lower than that within the ideal regime even for a wavelength as small as $\lambda = L/100$ confirming that dissipative effects persist. Noteworthy, we show in Sect. \ref{Appendix: growth rate parametric} that growth rates measured in simulations compare well with analytical predictions of the standard parametric instability. To summarize, although partially curtailed due to non-ideal MHD effects, dust clumping due to a parametric-like instability persists to some extent under specific conditions, namely high wave amplitude $\delta_B$ and low wavelength. 

In that regard, turbulence plays an essential role as it allows for the continuous and sustained production of Alfvén waves that span a large spectrum of wavelengths. Unstable Alfvén waves are rejuvenated over time and small wavelengths (inducing strong clumping) are reached, although the corresponding amplitudes of such modes cannot be arbitrary large due to energy decreasing with wavenumber \citep[if Kolmogorov spectrum, energy cascades as $E(k) \propto k^{-\frac{5}{3}}$, see ][]{Kim&Ryu2005}. Note however that because of the intermittent nature of the turbulence, energetic but localized events occur over very small scales in shocks, which is exactly where dust clumps form, as we shall see. Note however that our 1D simulations cannot produce all the important features of a 3D turbulent cascade. \newline In this work (when turbulent driving is on), the transverse-to-longitudinal ratio $B_\perp / B_\parallel$ is controlled by two parameters: the plasma parameter $\beta$ and the transverse Mach number $\mathcal{M_\perp}$. On one hand, increasing the former to a certain extent (in such a way as to keep a decent level of magnetization) reduces the background magnetic field $B_\parallel$ and thus increases the said ratio. On the other hand, a larger transverse Mach number implies higher transverse dust velocities which in turn (dust being the charged species) produce a stronger transverse magnetic field by bending and distorting $B_\parallel$. Another point has to be made regarding the impact of the plasma parameter. From theory and simulations of a single fluid found in the literature, $\beta$ is known to suppress the parametric instability for values $\beta \geq 1$ as thermal pressure begins to prevail over magnetic pressure, enabling sound waves to smooth out density fluctuations. However in this work, the magnetized fluid (dust) is pressureless and thus not subject to this process. We again stress that here, low values of $\beta$ are stabilizing while high values favor larger $B_\perp / B_\parallel$ and thus strong clumping of the dust. An informative illustration is provided in Fig. \ref{param instability cartoon}.   \newline
We now turn to turbulent simulations and look at Fig. \ref{Fields at different times} to follow the formation of dust clumps. This figure depicts various fields (dust and gas densities and velocities) at three different times, e.g. at $t=0.23$, $t=0.44$ and $t=0.82$. At first ($t=0.23$), dust and gas are tightly coupled, as seen in the nearly perfectly overlain profiles (in density and velocity). Shocks form in the gas and subsequently in the dust, giving rise to over-densities. In these high density regions, assuming the relevant Alfvén velocity to be that of the gas (valid in the strong coupling regime, see Sect. \ref{dispersion relation}), the said velocity is lowered (scales as $c_a \propto \rho^{-\frac{1}{2}}$) making it easier for the dust to be super-Alfvénic and consequently to distort magnetic field lines. Indeed, it is clear that the dust Alfvénic Mach numbers are significantly over unity there. In addition, we see that high dust Alfvénic Mach numbers spatially coincide with sharp magnetic pressure gradients, suggesting that field lines have been bent. Later, at $t=0.44$ and $t=0.82$, we see that prominent dust clumps have indeed formed in these regions. To summarize, turbulence in the gas aids in triggering what looks like a dust parametric instability in shocks (high density regions) where Alfvén velocity is lower and high transverse-to-longitudinal magnetic field ratio can arise as charged dust grains efficiently distort the background magnetic field. Long-lived high density dust clumps emerge owing to the absence of dust internal pressure (see Sect. \ref{CV test section}). Driven turbulent motions sustain the production of such clumps whose survival is imperilled by drag effects, especially magnetic drag which recouples both fluids. There is a competition between this instability and magnetic drag.

\subsection{Impact of parameters on dust and gas density fluctuations in turbulent simulations}
\label{parametric study}

\begin{table}
\caption{Range of values taken by the explored parameters.}
\begin{center}

\begin{tabular}{ |p{3cm}||p{4cm}|}
 \hline
 \multicolumn{2}{|c|}{Parameter exploration} \\
 \hline
 $\beta $&$[\boldsymbol{0.1},0.3,\boldsymbol{0.7},1]$\\

  $n_\mathrm{H} \left[\mathrm{cm^{-3}} \right] $&$[10^4,\boldsymbol{10^6},10^8]$\\
  
  $s_\mathrm{d} \left[\mu \mathrm{m} \right] $&$[0.1,1 ,\boldsymbol{10} ,100]$\\

  $\mathcal{M_\parallel} (\mathcal{M_\perp})$&$[0.12(0.17), 0.36(0.43)]$\\
  &$[\boldsymbol{0.56}\boldsymbol{(0.99)},0.92(1.91)]$\\
\hline

\end{tabular}
\tablefoot{In bold are the reference values used when a given parameter is not being varied. For every simulation: NX=4096 cells.}
\end{center}

\label{table1}
\end{table}

\begin{figure*}

\includegraphics[width=\textwidth]{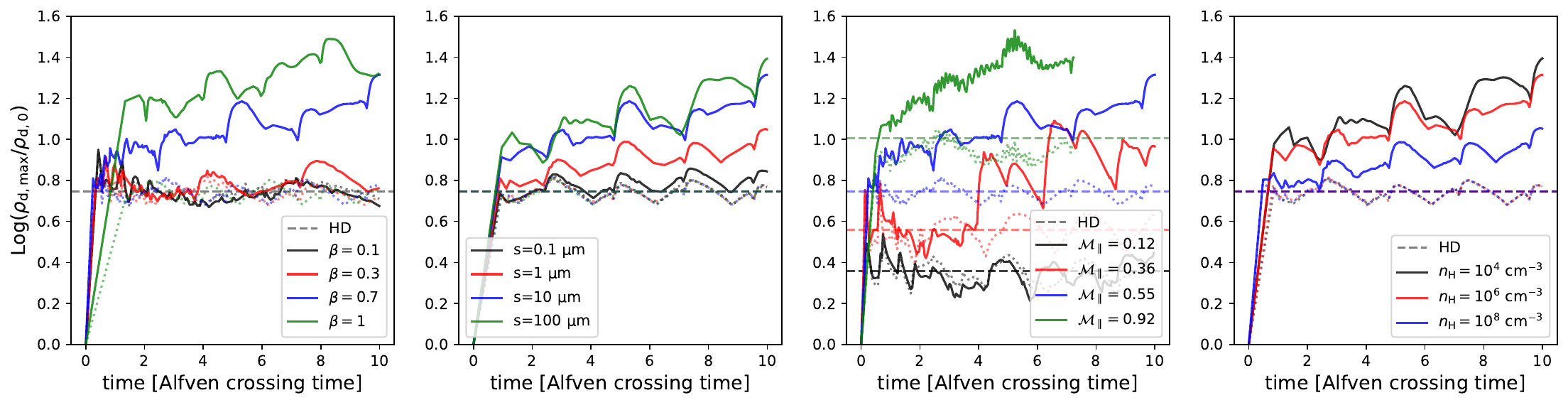}

\caption{Maximum dust density (solid lines) and gas density (dotted lines) fluctuations as a function of time for different values of the dust grain size $s_\mathrm{d}$, longitudinal Mach number (perpendicular one being varied too) $\mathcal{M}_\parallel$, gas initial density $n_\mathrm{H}$ and plasma parameter $\beta$. The dust density mean value for pure hydrodynamics simulations is displayed for reference as dashed lines. When not varied, $\beta=0.7$.}
\label{rhod vs time 0.7}
\end{figure*}

\begin{figure*}

\includegraphics[width=\textwidth]{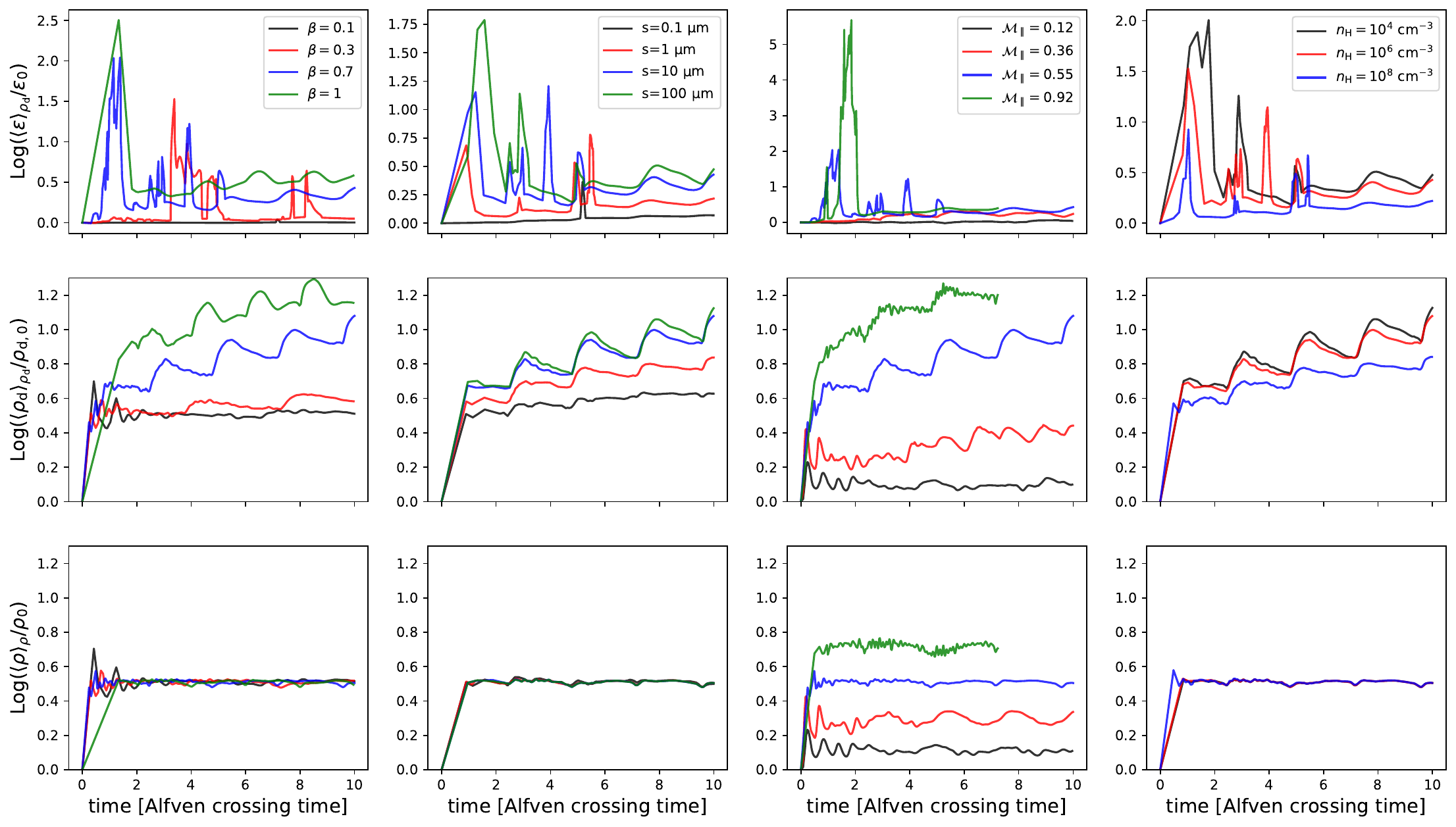}

\caption{Same as Fig. \ref{rhod vs time 0.7} but depicting  space averages (mass weighted) instead of maximum values. On the first row is displayed the dust-to-gas ratio (dust density weighted), on the second row the dust density (dust density weighted) and on the third row the gas density average (gas density weighted).}
\label{average vs time 0.7}

\end{figure*}

In this section, we drive turbulent motions and choose to vary four relevant parameters, namely the plasma parameter $\beta$, the gas initial numerical density $n_\mathrm{H} = \rho / \mu m_\mathrm{H}$, the dust grain size $s_\mathrm{d}$ (corresponding to a given $\mathrm{St}$) as well as the parallel and perpendicular Mach numbers $\mathcal{M_\parallel}$ and $\mathcal{M_\perp}$ (which we vary jointly). Table \ref{table1} shows the values taken for each parameter and in bold text the fiducial ones. \newline We point out that those are common astrophysical parameters that facilitate discussion and applications of our results to astrophysical environments. Nevertheless, based on Sect. \ref{section parametric instability}, it is worth keeping in mind  that their influence on dust clumping can be explained through their impact on two essential ingredients: the Stokes number of dust grains $\mathrm{St}$ and the transverse-to-longitudinal magnetic ratio $B_\perp / B_\parallel$. The latter controls the intensity of density fluctuations via the parametric-like instability, while the former encodes the degree of (hydrodynamical) coupling between gas and dust, crucial for the formation and survival of dust clumps.  \newline For clarity, we provide at the end of this section in Table \ref{table2} the equivalent $\mathrm{St}$ and maximum (time-averaged) $B_\perp / B_\parallel$ for each model (including those investigating the impact of a lower ion abundance; see Fig. \ref{beta ni rhod average}). Other useful data are gathered, including the time-averaged dust clumping fraction $C_{f,10}$ (defined as the
fraction of dust mass whose density has increased by a factor of more
than 10), along with the maximum and mean dust density fluctuation (spatially averaged first as in Eq. \ref{averaging formulae} and then time averaged over steady-state regime) and the dust clumping factor $\langle \rho_\mathrm{d}^2 \rangle/\langle \rho_\mathrm{d} \rangle^2$.

1. The plasma parameter $\beta$ takes values from 0.1 to 1, corresponding to the Alfvén velocity being 10 times the sound speed for the former (i.e., strong magnetization), and both being equal for the latter. For a gas density $n_\mathrm{H} = 10^4 \ \mathrm{cm^{-3}}$, values in the range $\beta = \left[0.1,0.3,0.7,1 \right]$ translate into the following range of magnetic field intensity $B_x = \left[41.6,24,15.7,13.2 \right] \ \mathrm{\mu G}$.  For $n_\mathrm{H} = 10^6 \ \mathrm{cm^{-3}}$, we get $B_x = \left[416,240,157,131 \right] \ \mathrm{\mu G}$, which are typical values observed in dense cores \citep[see][]{Crutcher2012,Pattle2023,Whitworth2025}. 

2. The initial numerical gas density ranges between $n_\mathrm{H} = 10^4 \ \mathrm{cm^{-3}}$ and $n_\mathrm{H} = 10^8 \ \mathrm{cm^{-3}}$. We keep $\beta$ fixed, which implies that $B_x$ increases as the square root of $n_\mathrm{H}$ when varying it, as seen in Eq. (\ref{plasma beta}). This is a reasonable assumption since the magnetic field is expected to behave roughly this way as collapse proceeds \citep[][]{Li2011,Pattle2023}. Therefore, when increasing $n_\mathrm{H}$, we probe different stages of gravitational collapse of dense cores.

3. Dust grains have sizes ranging from $0.1 \ \mathrm{\mu m}$ to $100 \ \mathrm{\mu m}$. The lowest value lies within the size range of the Mathis-Rumple-Nordsiek distribution \citep{Mathis1977} observed in the ISM. $1 \ \mathrm{\mu m}$ corresponds to a few times the upper value of the MRN distribution, i.e., to dust grains that have undergone some growth beforehand with respect to ISM dust. We also explore larger sizes to gain insight. The size range translates into a Stokes number range: $\mathrm{St} = \left[10^{-4},10^{-3},10^{-2},10^{-1}\right]$.

4. The parallel Mach number values we explore lie between $\mathcal{M}_\parallel = 0.12$ and $\mathcal{M}_\parallel = 0.92$ while the perpendicular Mach number covers larger values, ranging from $\mathcal{M}_\perp = 0.17$ to $\mathcal{M}_\perp = 1.91$. This choice of range is justified by turbulence in dense cores being observed as subsonic \citep{Barranco1998,Choudhury2021}. Again, it should be kept in mind that we drive 1D turbulence, which is inherently distinct from 3D turbulence. \\

We find that the plasma parameter $\beta$ is the most influential. For a value as low as $\beta
 = 0.1$, the other parameters are very little impactful, and the associated plots are supplemented in the appendix, Sect. \ref{Dust clumping beta0.1}, while we provide in the present section the results for $\beta = 0.7$ for which dust clumping is significant. Every simulation has been run with a resolution of 4096 cells and we note that convergence is not reached regarding dust density. The discussion of this point is delayed to Sect. \ref{Section CV and caveats}.
 
\begin{figure*}
\centering

    \includegraphics[width=\textwidth]{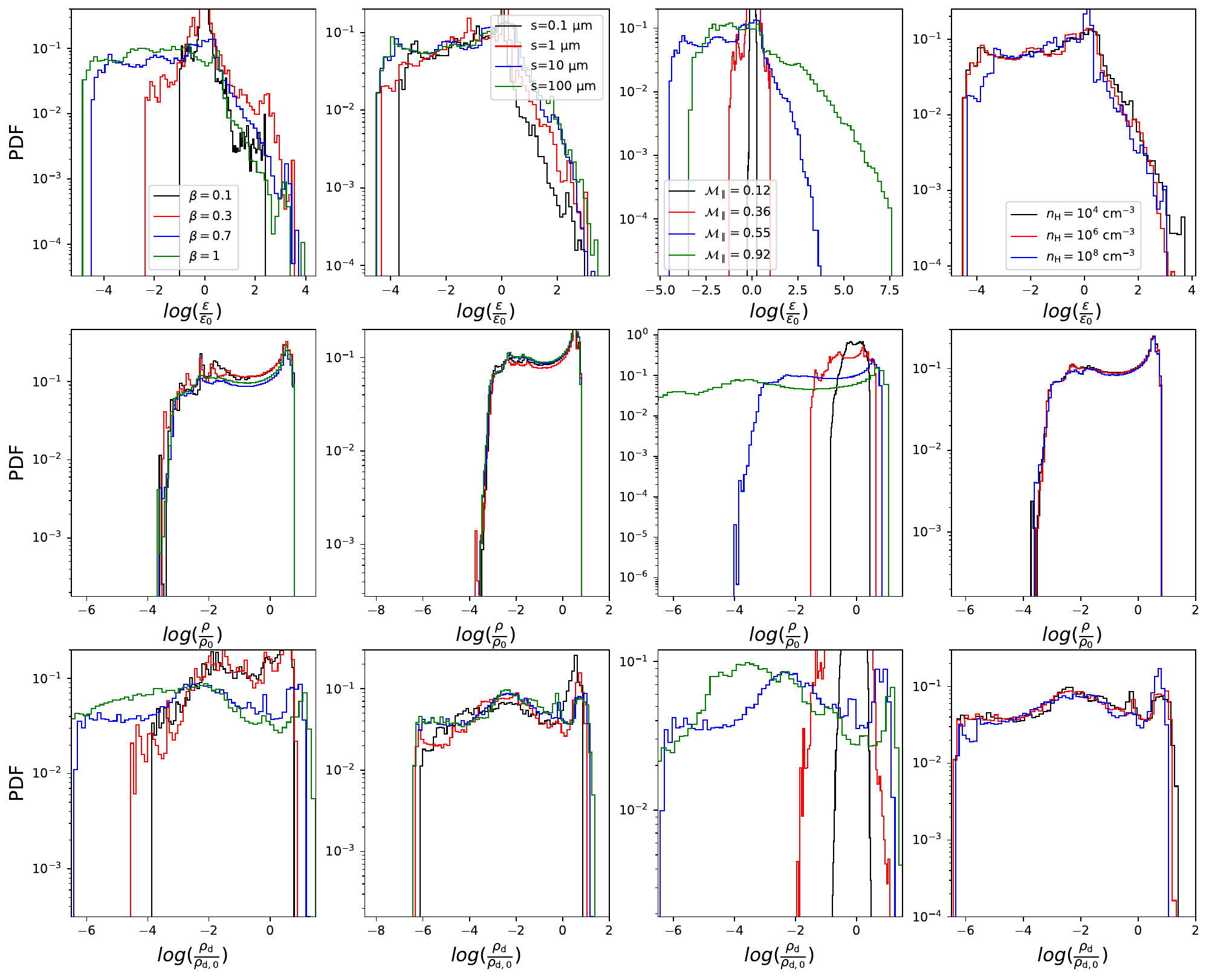}
    \caption{Time-averaged (over driven turbulence steady-sate regime) probability distribution functions (PDF) of dust-to-gas ratio fluctuations (first row), gas density fluctuations (second row) and dust density fluctuations (third row). }
    \label{1DHist}
\end{figure*}

 Figure \ref{rhod vs time 0.7} depicts the maximum dust density (solid lines) and gas density (dotted lines) fluctuations as a function of time for different values of the dust grain size, longitudinal Mach number (perpendicular also varied), gas initial density and plasma parameter. The dust density mean value for pure hydrodynamics simulations is displayed for reference as dashed lines. Figure \ref{average vs time 0.7} displays the fluctuations of the same quantities, but space-averaged (mass weighted) as follows:
 \begin{equation}
     \langle a \rangle_{a} = \int_0^L a.a dx/\int_0^L adx = \int_0^L a^2 dx/\int_0^L adx,
\label{averaging formulae}
 \end{equation} 
where $a$ is either the dust density or gas density. Both are self-weighted, allowing us (for instance if $a = \rho_\mathrm{d}$) to capture regions of dust enrichment. The dust-to-gas ratio average is however weighted by the dust density. Besides, in order to discard high dust-to-gas ratios induced by severe gas depletion (where the dust could be depleted as well but to a lesser extent), we apply a filter and retrieve the dust-to-gas ratio only in cells where $\rho_\mathrm{d} > \rho_\mathrm{d,0}$. Indeed, this corresponds to regions of dust concentration, which is what is of interest when it comes to dust growth. Time is measured in Alfvén crossing time, defined as the time needed for an Alfvén wave to cross the box of length $L$, i.e., as $t_\mathrm{a} = L/c_\mathrm{a}$. \newline
Figure \ref{1DHist} shows the time-averaged (over driven turbulence steady-sate regime) probability distribution functions (PDF) of dust-to-gas ratio fluctuations (first row), gas density fluctuations (second row) and dust density fluctuations (third row). \newline

\subsubsection{Impact of plasma parameter}
 We first explore the impact of the plasma parameter $\beta$ which, as mentioned above, is found to be the most influential one. Looking at Fig. \ref{rhod vs time 0.7}, we notice a high sensitivity of the maximum dust density to this parameter. Indeed, at $\beta = 0.1$, the dust maximum density is enhanced by a factor of $10^{0.8} = 6.31$ in the steady-state regime relative to its initial value. For larger values of $\beta$, we see a stronger clumping of dust whose density is enhanced by a factor of over $10^{1.2} = 15.85$ ($\beta=1$). In contrast, gas density fluctuations do not vary as we increase $\beta$ and remain near a value slightly below $10^{0.8}$. This is also the average value of both gas and dust density fluctuations in pure hydrodynamics simulations (pictured as a dashed line) suggesting that neutral gas concentration is controlled solely by hydrodynamical turbulent motions while dust concentration is driven by magnetic effects. This behavior is confirmed in Fig. \ref{average vs time 0.7}. \newline
 As explained in detail in Sect. \ref{section parametric instability}, the process responsible for dust concentration is most likely the parametric instability, which develops more efficiently for large transverse-to-longitudinal magnetic field ratios $B_\perp/B_\parallel$. As we increase $\beta$, we reduce the longitudinal magnetic field component $B_\parallel = B_x$ and thus increase the said ratio. When a low value of $\beta = 0.1$ is considered, magnetic effects do not contribute to dust concentration. Instead, dust and gas remain tightly coupled as ions force neutral gas molecules to move along magnetic field lines (magnetic drag, see Eq. \ref{final dust mom eq}) and only modest density fluctuations are observed (see Fig. \ref{rhodvstime0.1}). When $\beta$ increases and magnetic effects prevail, sharper dust clumps are allowed to form and decouple from the gas, resulting in a broader spectrum of dust-to-gas ratios, as seen in Fig. \ref{1DHist}. \newline We show in the Appendix (Fig . \ref{B_perp over B_par vs beta}) the evolution of the maximum $B_\perp/B_\parallel$ as a function of time in simulations with turbulence for different values of $\beta$. We see clearly that stronger $B_\perp/B_\parallel$ are achieved for larger values of $\beta$.

\subsubsection{Impact of grain size}
\label{Section impact of grain size}

We investigate here the impact of grain size. As a reminder, when not varied, the plasma parameter takes its fiducial value $\beta=0.7$ allowing dust density fluctuations to be enhanced. As readily seen in Fig. \ref{rhod vs time 0.7}, larger grains are prone to stronger density fluctuations. For small grains of $s_\mathrm{d} = 0.1 \ \mathrm{\mu m}$ in size, a modest level of dust concentration is reached which is equal to that of gas and comparable to what we get from pure hydrodynamics simulations. Both fluid are very well coupled and magnetic forces cannot produce sharp dust clumps. However, for marginally coupled dust grains ($s_\mathrm{d} \geq 10 \ \mathrm{\mu m}$, that is $\mathrm{St} \geq 10^{-2}$), an increase of more than a factor of $10$ in dust density is easily attained, which continues to increase even after $10$ Alfvén crossing times.
Even relatively small particles of size $s_\mathrm{d} \geq 1 \ \mathrm{\mu m}$ experience remarkable density enhancements, close to a factor of $10$. Interestingly, the gain in clumping efficiency with size seems to approach saturation as we go from $s_\mathrm{d} = 10 \ \mathrm{\mu m}$ to $s_\mathrm{d} = 100 \ \mathrm{\mu m}$, meaning that the largest grains reach a state of full decoupling from the gas, which thus ceases to influence their dynamics. Noteworthy, dust-to-gas ratio fluctuations also increase with size, reaching values above $100$ and as high as $1000$ for $s_\mathrm{d} \geq 100 \ \mathrm{\mu m}$ (see Fig. \ref{average vs time 0.7}).

This dependency of dust concentration on grain size has interesting consequences in terms of dust coagulation. As discussed in Sect. \ref{Section coag instability}, this could lead to a runaway dust growth.

\subsubsection{Impact of Mach number}

We now turn our attention to the influence of the Mach number. Note that computational cost of simulations sharply increases with this parameter, therefore we chose not not to integrate the run for $\mathcal{M}_\parallel = 0.92$ beyond $4$ Alfvén crossing times. \newline
First of all, we notice that unlike the previous parameters, varying the Mach number does affect gas density fluctuations. Indeed, increasing the turbulent Mach number leads to stronger fluctuations in the gas density, reaching higher maximum values and strikingly lower minimum values, i.e., a much more inhomogeneous distribution of matter. Figure \ref{1DHist} clearly shows that gas density fluctuations as low as $10^{-6}$ are not uncommon in regions of gas depletion. The maximum gas density increases by a factor of $10^{0.4} = 2.5$ for $\mathcal{M}_\parallel = 0.12$ to up to $10$ for $\mathcal{M}_\parallel = 0.92$ (third column, third row). Again, those values match those from hydrodynamics simulations, showing that gas clumping is due to non magnetic turbulent effects only.  \newline
As for the dust, we observe a level of clumping similar to that of gas in the lowest Mach number run ($\mathcal{M}_\parallel = 0.12$), implying that both fluids are well coupled and concentrate jointly. However, as the Mach number increases, a substantial departure arises and dust is allowed to exhibit stronger density enhancements, by a factor of about $10$ for $\mathcal{M}_\parallel = 0.56$ and over $10^{1.4} = 25.1$ for $\mathcal{M}_\parallel = 0.92$ (third column, second row). Here again, this behavior should be attributed to magnetic effects, keeping in mind that the transverse Mach number $\mathcal{M}_\perp$ is varied along with its longitudinal counterpart. As explained in Sect. \ref{section parametric instability} and discussed in Sect. \ref{turb as a good candidate}, the transverse-to-longitudinal magnetic ratio $B_\perp / B_\parallel$ (and thus the parametric instability) is controlled by $\mathcal{M}_\perp$ (see Table \ref{table2}).  

Finally, while it is striking that the maximum dust-to-gas ratio significantly increases with Mach number, the minimum value does not drop as low for $\mathcal{M}_\parallel = 0.92$ as for $\mathcal{M}_\parallel = 0.56$, meaning that in supersonic regime, the gas is less inclined to concentrate in regions devoid of dust.

\subsubsection{Impact of background gas density}

Finally, we explore the impact of the initial background gas density $n_\mathrm{H}$. Although local gas density fluctuations remain unaffected, dust clumping appears to be reduced in denser environments. Indeed, while maximum fluctuations by a factor of $10$ are easily reached at $n_\mathrm{H} = 10^{4} \ \mathrm{cm^{-3}}$ and $n_\mathrm{H} = 10^{6} \ \mathrm{cm^{-3}}$, they remain below this threshold at $n_\mathrm{H} = 10^{8} \ \mathrm{cm^{-3}}$ (see Fig. \ref{rhod vs time 0.7}, fourth panel). This is caused by the competition of two different effects, one inhibiting dust concentration and the other promoting it. First, a higher background gas density implies a lower Stokes number $\mathrm{St}$ (a lower stopping time $t_\mathrm{s,d}$, see Eq. \ref{ts}) for a given dust grain size, that is, a better coupling between dust and gas, preventing efficient dust clumping. Second, we expect on average that the ion numerical density relative to dust density is lower in denser environments since $\rho_i \propto \sqrt{n_\mathrm{H}}$ (see Eq. \ref{Shu}) and $\rho_\mathrm{d,0} = \epsilon \rho_0 \propto n_\mathrm{H}$. As discussed in Sect. \ref{section non-ideal effects and chemical network}, the impact of magnetic drag due to ions should be reduced compared to that of hydrodynamical drag, allowing for an easier dust clumping. However, this effect is negligible with respect to the influence of a lower $\mathrm{St}$, and the overall tendency is a reduction in dust density fluctuations. \newline Note that, as per Eq. (\ref{plasma beta}), a higher density with a fixed $\beta$ yields a stronger longitudinal magnetic field $B_\parallel = B_x$, and one could think that for a given level of turbulence ($\mathcal{M}_\perp$), higher transverse-to-longitudinal magnetic ratio $B_\perp / B_\parallel$ could be reached. However, it is not the case, because a fixed $\beta$ implies that the initial gas density $n_\mathrm{H}$ and the background $B_x$ are varied jointly in order to keep the same Alfvén  velocity $c_\mathrm{a}$. As a consequence, similar dust Alfvénic Mach numbers are achieved, giving rise to very similar $B_\perp / B_\parallel$ (see Table \ref{table2}). Thus, it is only the decrease in $\mathrm{St}$ that leads to a less efficient clumping of the dust at higher $n_\mathrm{H}$. 

In Fig. \ref{1DHist}, we see that the dust-to-gas ratio PDF is slightly narrower at $n_\mathrm{H} = 10^{8} \ \mathrm{cm^{-3}}$ due to tighter coupling between gas and dust.

\begin{table*}
\caption{Stokes number $\mathrm{St}$, time-averaged maximum transverse-to-longitudinal magnetic ratio $B_\perp / B_\parallel$, dust clumping fraction $C_{f,10}$, maximum and mean dust density fluctuation, and dust clumping factor $\langle \rho_\mathrm{d}^2 \rangle/\langle \rho_\mathrm{d} \rangle^2$ for the different models.}
\begin{center}
\begin{tabular}{|p{2cm}|p{1cm}|p{1cm}|p{1cm}|p{1.5cm}|p{1cm}|p{1.5cm}|}
 \hline
\makebox[2cm]{Model}   
& \makebox[1cm]{\centering $\mathrm{St}$}
& \makebox[1cm]{\centering $B_\perp / B_\parallel$}
& \makebox[1cm]{\centering $C_{f,10}$} & \makebox[1.5cm]{\centering $\rho_\mathrm{d,max}/\rho_\mathrm{d,0}$} & \makebox[1cm]
{\centering $\langle \rho_\mathrm{d} \rangle/\rho_\mathrm{d,0}$} 
& \makebox[1.5cm]
{\centering $\langle \rho_\mathrm{d}^2 \rangle/\langle \rho_\mathrm{d} \rangle^2$}\\
 \hline
 $\beta 0.1 $&$10^{-2}$&$0.05$&$0$&$5.59$&$3.20$&$1.22$\\
 $\beta 0.3 $&$10^{-2}$&$0.15$&$0$&$6.57$&$3.82$&$1.29$\\
 $\beta 0.7 $&$10^{-2}$&$0.37$&$0.35$&$12.8$&$7.76$&$2.0$\\
 $\beta 1 $&$10^{-2}$&$0.52$&$0.69$&$20.5$&$13.2$&$2.21$\\

  $n_\mathrm{H} 10^4$&$10^{-1}$&$0.36$&$0.41$&$15.2$&$8.35$&$2.26$\\
  $n_\mathrm{H} 10^6$&$10^{-2}$&$0.36$&$0.35$&$12.8$&$7.76$&$2.0$\\
  $n_\mathrm{H} 10^8$&$10^{-3}$&$0.38$&$0.01$&$8.21$&$5.45$&$1.52$\\

  $s_\mathrm{d} 0.1  $&$10^{-4}$&$0.38$&$0$&$6.23$&$3.95$&$1.27$\\
  $s_\mathrm{d} 1  $&$10^{-3}$&$0.37$&$0.02$&$8.44$&$5.56$&$1.53$\\
  $s_\mathrm{d} 10  $&$10^{-2}$&$0.36$&$0.34$&$12.8$&$7.71$&$2.0$\\
  $s_\mathrm{d} 100  $&$10^{-1}$&$0.36$&$0.4$&$14.9$&$8.18$&$2.25$\\

  $\mathcal{M_\parallel} 0.12$&$10^{-2}$&$0.11$&$0$&$2.22$&$1.25$&$1.17$\\
  $\mathcal{M_\parallel} 0.36$&$10^{-2}$&$0.25$&$0.03$&$9.13$&$2.68$&$1.59$\\
   $\mathcal{M_\parallel} 0.56$&$10^{-2}$&$0.37$&$0.35$&$12.4$&$7.41$&$1.96$\\
   $\mathcal{M_\parallel} 0.92$&$10^{-2}$&$0.45$&$0.76$&$21.9$&$14.4$&$2.21$\\
 
  $\beta 1 s_\mathrm{d} 0.1  $&$10^{-4}$&$0.55$&$0$&$6.51$&$4.27$&$1.31$\\
  $\beta 1 s_\mathrm{d} 1  $&$10^{-3}$&$0.54$&$0.22$&$11.5$&$7.71$&$1.93$\\
  $\beta 1 s_\mathrm{d} 100  $&$10^{-1}$&$0.51$&$0.72$&$26.8$&$15.6$&$3.85$\\

  $\beta 1 n_\mathrm{i}/10 s_\mathrm{d} 0.1  $&$10^{-4}$&$0.52$&$0$&$6.53$&$4.28$&$1.31$\\
  $\beta 1 n_\mathrm{i}/10 s_\mathrm{d} 1  $&$10^{-3}$&$0.49$&$0.35$&$12.9$&$8.25$&$2.26$\\
  $\beta 1 n_\mathrm{i}/10 s_\mathrm{d} 10  $&$10^{-2}$&$0.43$&$0.76$&$58.7$&$24.2$&$8.25$\\
  $\beta 1 n_\mathrm{i}/10 s_\mathrm{d} 100  $&$10^{-1}$&$0.38$&$0.95$&$515.6$&$328.9$&$4240.4$\\
\hline

\end{tabular}
\tablefoot{When not indicated, parameter values are implicitly assumed to be the fiducial ones (see Table \ref{table1}). The clumping fraction is defined as the fraction of dust mass whose density has increased by a factor of more than 10 ($\rho_\mathrm{d}/\rho_\mathrm{d,0} \geq 10$). The spatially averaged dust density fluctuation in computed as per Eq. (\ref{averaging formulae}). Those values are then averaged over time (steady-state regime).}
\end{center}
\label{table2}
\end{table*}

\section{Discussion}
\label{discussion}

\subsection{Non-ideal effects and chemical network}
\label{section non-ideal effects and chemical network}

As seen in Fig. \ref{idealvsnonideal plot} in Sect. \ref{Ideal vs non-ideal} (see also Fig. \ref{param instability ideal vs non-ideal}), dust clumping is curtailed to a certain extent when considering non-ideal MHD effects. We thus see that there is a need to consider microphysical processes along with an accurate description of MHD equations to predict relevant physical quantities such as density fluctuations. First, in simulations, ions cannot be considered as main charge carriers in dense cores, and therefore gas should be considered neutral \citep[contrary to more diffuse environments such as molecular clouds, see][]{Lee&Hopkins2017,Moseley2023}. Second, the present work demonstrates that assuming a perfect coupling between dust particles and magnetic field lines (Sect. \ref{dusty ideal MHD}) is inaccurate. Non-ideal MHD effects do make a difference. \newline 
We found the discrepancies between dusty ideal MHD and dusty non-ideal MHD models to be primarily attributed to the additional magnetic drag term induced by the presence of ions in the dust momentum equation (Eq. \ref{final dust mom eq}) that we rewrite $\frac{e n_\mathrm{d} Z_\mathrm{d} B}{\Gamma_\mathrm{i} c}(\vec{v}-\vec{v}_\mathrm{d}) = \frac{\rho_i}{t_\mathrm{s,i}}(\vec{v}-\vec{v}_\mathrm{d})$. It is instructive to compare this magnetic drag term with the hydrodynamical one, i.e. to compare $\frac{\rho_i}{t_\mathrm{s,i}}$ with $\frac{\rho_\mathrm{d}}{t_\mathrm{s,d}}$. We first compute stopping times. Using the fiducial background gas density $\rho = 3.9 \times 10^{-18} \ \gcc$ (corresponding to $n_\mathrm{H} = 10^6 \ \mathrm{cm^{-3}}$) and an ion ($\mathrm{HCO^+}$) gas collision rate $\gamma_\mathrm{i,g} = 3.5 \times 10^{13} \ \mathrm{cm^{-3} \ g^{-1} \ s^{-1}}$ given in \cite{Pinto2008}, we find $t_\mathrm{s,i} = \left(\rho \gamma_\mathrm{i,g} \right)^{-1} \simeq 10^4 \ \mathrm{s}$. From Eq. (\ref{ts}) for $s_\mathrm{d} = 100 \ \mathrm{\mu m}$ ($\mathrm{St} = 0.1$) we compute the dust stopping time $t_\mathrm{s,d} \simeq 2 \times 10^{11} \ \mathrm{s}$. In terms of orders of magnitude, it yields:
\begin{align}
     \frac{\rho_\mathrm{d}}{t_\mathrm{s,d}}\simeq\frac{10^{-20}}{10^{11}} = 10^{-31} \ \mathrm{g \ cm^{-3} \ s^{-1}},  \nonumber \\   \frac{\rho_i}{t_\mathrm{s,i}}\simeq\frac{10^{-25}}{10^{4}} = 10^{-29} \ \mathrm{g \ cm^{-3} \ s^{-1}},
\end{align}
where $\rho_i$ was computed from Eq. (\ref{Shu}) (see also Fig. \ref{Shu vs Marchand}). The higher mass density of dust grains does not compensate for the enormous gap in stopping times. Magnetic drag dominates by a factor of $\simeq 100$. Since $t_\mathrm{s,d} \propto s_\mathrm{d}$, the two drag terms become comparable if we decrease the dust size by two orders of magnitude, i.e., for $s_\mathrm{d} \leq 1 \ \mathrm{\mu m}$.

Based on the previous calculations, we see that magnetic drag prevails under typical conditions. As long as it does, dust grains remain very well coupled to the gas and cannot concentrate independently. Unlike the dust stopping time, the ion stopping time is not a function of grain size. Consequently, in this regime, larger grains do not decouple from the gas as we usually expect from hydrodynamical drag. This explains the modest dust density fluctuations observed in Fig. \ref{rhodvstime0.1} (where $\beta = 0.1$) and in particular why we do not see a dependency of the dust density fluctuations on grain size. This behavior is supported by the impact of Mach number, where we see that regardless of its value, both dust (solid lines) and gas (dotted lines) clump by the same amount (they are tightly coupled by magnetic drag) and reach the same values as those obtained in pure hydrodynamical simulations (dashed lines). \newline
There is a competition between the parametric-like instability identified in Sect. \ref{section parametric instability} and magnetic drag. The former favors dust concentration, while the latter inhibits it. As already mentioned, we need to increase the transverse-to-longitudinal magnetic field ratio $B_\perp / B_\parallel$ to enhance instability and hence dust clumping. To do so, we can either increase the plasma parameter $\beta$ (i.e., decrease $B_x$) or increase the transverse Mach number $\mathcal{M}_\perp$ (see Fig. \ref{param instability cartoon}). Indeed, the induction equation tells us that the background longitudinal magnetic field $B_x$ is distorted, producing transverse components $B_y$ and $B_z$ if the main charge carrier (here the dust) is allowed to move in transverse directions. On the other hand, to reduce magnetic drag, one could play with the ion numerical density $n_\mathrm{i}$. Decreasing it would allow dust grains to decouple more easily from the gas and clump.
This point leads us to discuss our choice of chemical network. \newline
Although our approach (see Sect. \ref{chemical network}) offers simplicity, it is not fully realistic. First of all, there is no dependency of the total numerical density of charge $n_\mathrm{d}Z_\mathrm{d}$ of the dust fluid on grain size. Although it is expected to increase with dust size \citep{Draine87}, the exact charge scaling with size is not straightforward when considering various reactions such as dust-electron capture, dust-ion capture, ion-electron recombination on grain surface or even grain-grain reactions \citep[e.g. charge transfer, see ][]{Waitukaitis2014}. Second, a more sophisticated chemical network including a dust size distribution would lead to a lower ion abundance owing to very small grains offering a large total cross section for ion-electron recombination on their surface. Upon comparison with the chemical network of \cite{Marchand2021} (see Fig. \ref{Shu vs Marchand}) where a dust size distribution (MRN like) is considered, ions should be a few times less than predicted by our chemical network at about $n_\mathrm{H} = 10^6 \ \mathrm{cm^{-3}}$, leading to a weaker magnetic drag. The departure from one model to the other is reduced at lower density but is more severe at higher density. Note that abundances also depend on the cosmic ray ionization rate $\zeta$.

\subsection{Dust growth in protostellar envelopes}

Recent observational studies have revealed low dust emissivity index ($\beta < 1$)\footnote{Not to be mistaken for the plasma parameter.} in protostellar envelopes (at radii from $100 \ \mathrm{AU}$ to $500 \ \mathrm{AU}$), suggesting the presence of dust grains ranging between $100 \ \mathrm{\mu m}$ and $1 \ \mathrm{mm}$ in size \citep{Galametz2019,Cacciapuoti2023,Cacciapuoti2025}, supported by polarization measurements \citep{Valdivia2019,Sadavoy2019}. Indeed, the emissivity index $\beta$ is anti-correlated with the maximum grain size of the distribution where values $\beta \geq 1.5$ are consistent with dust sizes $s_\mathrm{d} \leq 10 \ \mathrm{\mu m}$, while $\beta \leq 1$ indicate the presence of $s_\mathrm{d} \sim 100 \ \mathrm{\mu m}$ dust grains. Although other considerations could explain this particularly low value \citep[effects of temperature and optical depth, contamination by synchrotron radiation, free-free emission and spinning dust emission, and in particular grain structure and composition effects, see][]{Carpine2025}, the actual presence of significantly grown dust grains is more than plausible \citep{Demyk2017,Ysard2019}.

Two (not mutually exclusive) scenarios have been proposed to explain these observations, namely the possibility for the grains to grow in-situ by means of coagulation, and the ejection of already grown grains via winds and outflows from the inner regions of the protoplanetary disk into the inner/intermediate regions of the envelope \citep[see][for the latter]{Wong2016,Tsukamoto2021}. This mechanism depends on some criteria including the mass of the central stellar embryo, the gas velocity, the opening angle of the outflow and the mass loss rate associated. The mass of the central nascent star needs to be sufficiently low, and the outflow mass ejection rate high enough to ensure an efficient transportation of the grains. 

When it comes to in-situ grain growth, current coagulation models face difficulties to form such large grains in a reasonable amount of time under standard assumptions \citep[see][]{Silsbee2022}. However, we stress here that this scenario cannot be ruled out yet as it seems possible to form millimeter grains in-situ provided that the dust is concentrated locally. In this section, we discuss the clumping measurements made in our simulations and the likelihood of this scenario.

\subsubsection{Is turbulent clumping a good candidate?}
\label{turb as a good candidate}

\begin{figure*}[hbt]

    \sidecaption
    \includegraphics[width=12cm]{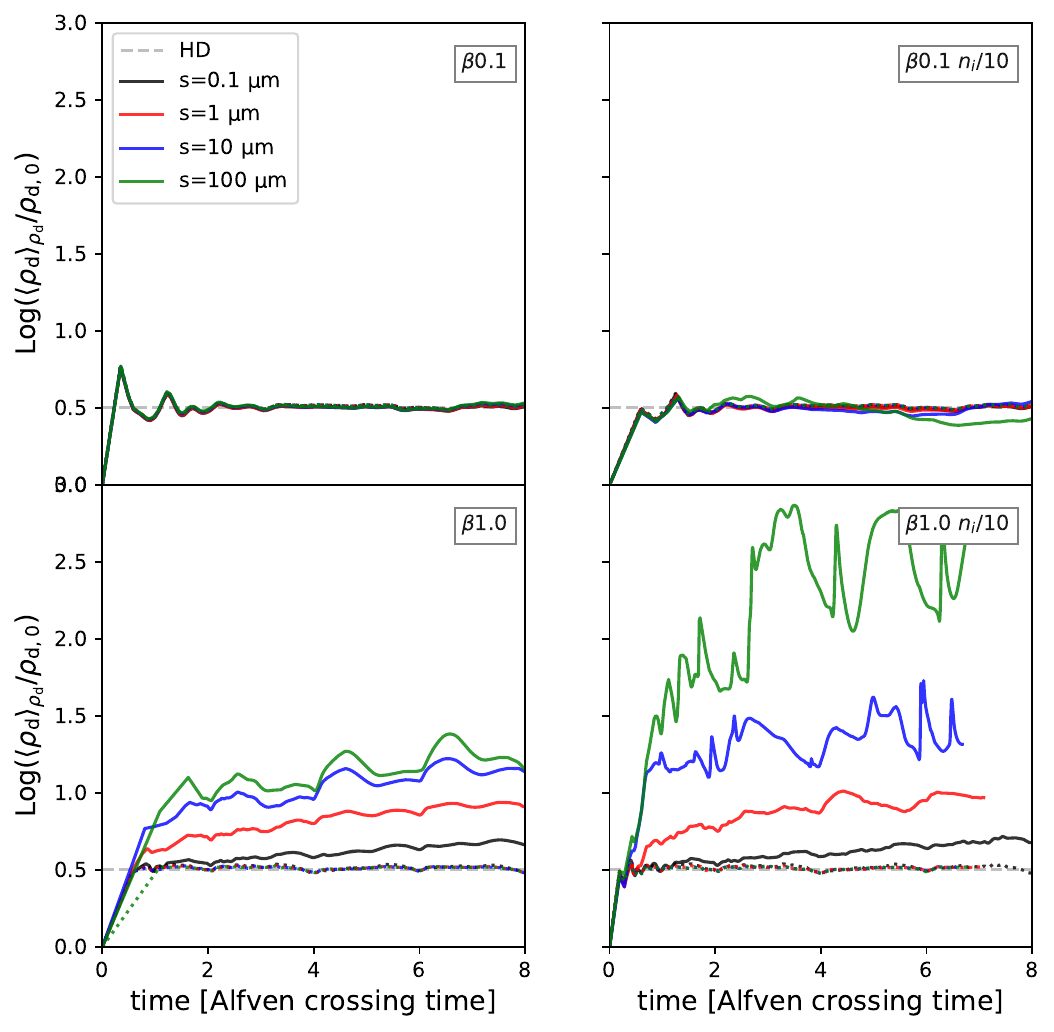} 
    \caption{Average (dust density weighted) dust density (solid lines) and average (gas density weighted) gas density (dotted lines) fluctuations as a function of time for different values of the dust grain size $s_\mathrm{d}$. Upper left panel: $\beta=0.1$. Lower left panel: $\beta=1$. Second column: same but with ten times less ions.  The dust density mean value for pure hydrodynamics simulations is displayed for reference as dashed lines.}
    
    \label{beta ni rhod average}
\end{figure*}

We saw that the two main ingredients when it comes to dust clumping are the transverse-to-longitudinal magnetic field ratio $B_\perp / B_\parallel$ and the ion numerical density $n_\mathrm{i}$. 
One way to increase the former, is by increasing the transverse Mach number $\mathcal{M}_\perp$. The corresponding effect is readily seen in Fig. \ref{rhod vs time 0.7}. Although dense core environments are expected to be sub-/transonic \citep{Barranco1998,Choudhury2021}, certain mechanisms could generate supersonic turbulence \citep[][explore the generation and amplification of turbulence in collapsing environments (gravo-turbulence)]{Hennebelle2021,Higashi2021}.
Studies show that a significant fraction of the velocity dispersion appears as a systematic drift that cannot be interpreted as turbulence. However, systematic drift would lead to instabilities \citep[resonant drag instabilities, ][]{Hendrix2014,Squire2018} favoring dust concentration. The rest is in the form of random velocity fluctuations which can be seen as supersonic turbulence on smaller scales. In massive dense cores, turbulence is expected to reach supersonic regime, although this claim has recently been questioned \citep[see][]{Wang2024}. Figure \ref{rhod vs time 0.7} shows that for $\mathcal{M}_\parallel = 0.56$ ($\mathcal{M}_\perp = 0.99$) and $\beta=0.7$, an increase by a factor of 0.8-1.0 in the average dust density is expected, which would allow for rapid grain growth. Going supersonic leads to even stronger dust clumping.  

Another way to increase the transverse-to-longitudinal magnetic field ratio for a given Mach number, is by increasing the plasma parameter $\beta$. Although too weak a magnetic field would suppress the clumping observed in our simulations ($\beta \longrightarrow \infty$ implies $\vec{B} \longrightarrow \vec{0}$), reducing the magnetization of the background magnetic field to a certain extent does favor dust clumping via parametric instability (Sect. \ref{section parametric instability}). Indeed, for a given turbulent driving (i.e., a given transversal Mach number $\mathcal{M}_\perp$), the level of dust concentration is very sensitive to $\beta$ as it controls the transverse-to-longitudinal magnetic ratio $B_\perp / B_\parallel$ (see Table \ref{table2} and Fig. \ref{B_perp over B_par vs beta}). Figure \ref{rhodvstime0.1} indicates that the average dust density increases only slightly (by a factor of $\sqrt{10}$) relative to initial value for $\beta=0.1$ compared with a magnetization seven times lower ($\beta=0.7$), which yields a dust density enhancement of more than a factor of ten (Fig. \ref{rhod vs time 0.7}). Although $\beta=0.1$ is a typical value for protostellar cores, a distribution of values is measured observationally \citep[][]{Crutcher2012,Pattle2023} and found in simulations \citep[][]{Hennebelle2011,Tu2022}. Cores threaded by weaker background magnetic fields would be more sensitive to this mechanism and prone to significant dust clumping.
Nevertheless, one should keep in mind the timescales of interest. Our results suggest that significant clumping is achieved after about 1 to 2 Alfvén crossing times. When $\beta=1$, by definition, the Alfvén crossing time corresponds to the sonic crossing time, and since we defined our box length $L$ as the Jeans length, such a time is equal to the free-fall time. Fortunately, protostellar envelopes should persist over a few free-fall times owing to other support mechanisms, thus offering enough time for dust grains to grow significantly. However, this conclusion should not hold for larger values of $\beta$. 

We explained in Sect. \ref{section non-ideal effects and chemical network} that fewer ions would allow dust grains to decouple from the gas and thus accumulate more efficiently. Based on the associated discussion, we explore here the impact of considering a 10 times lower ion abundance: $n_\mathrm{i} \rightarrow n_\mathrm{i}/10$. This is readily seen in Fig. \ref{beta ni rhod average} which depicts a striking soaring in the dust density when combining a lower magnetization ($\beta=1$) and ten times less ions. Micron-sized ($s_\mathrm{d} = 1 \ \mathrm{\mu m}$) grains display a increase by a factor of 10 in density, while a factor of about $30$ is reached for grains ten times bigger ($s_\mathrm{d}=10 \ \mathrm{\mu m}$). Dust density fluctuations climb as high as $10^{2.5} = 300$ for grains of size $s_\mathrm{d}=100 \ \mathrm{\mu m}$. Considering ion numerical densities ten times lower than those predicted by our chemical network is not unreasonable since it matches the values obtained with a much more sophisticated chemical network including a distribution of grain size (see Sect. \ref{Chemical network comparison}), where very small grains would efficiently recombine ions and electrons on their surface. Despite a swift coagulation owing to ambipolar drift, the small grain population is expected to persist at low densities in dense cores due to electrostatic repulsion \citep[investigated in protoplanetary disks in][]{Akimkin2023} and grain-grain erosion \citep{Vallucci2024}. Note however, that if sufficiently well coupled to the magnetic field, very small grains could behave as ions to some extent and maintain a certain level of magnetic drag between gas and larger grains, making efficient clumping more difficult to achieve. On the other hand, we should keep in mind that very small dust grains can carry only a single electric charge (just as the ions we consider here) but are substantially more massive than ions. This means that their (small dust grains) Hall factor would be higher and thus the subsequent induced magnetic drag lower (according to Eq. \ref{final dust mom eq}) than that induced by ions in the present work.    
Simulations with a distribution of dust sizes and a more realistic chemical network are needed to address this point, but require substantial developments that are beyond the scope of this work.

In addition, a different ionization rate induced by cosmic rays $\zeta$ would change the abundance of ions as well. In Eq. (\ref{Shu}), we assumed $\zeta = 1 \times 10^{-17} \ \mathrm{s^{-1}}$. However, this parameter is not tightly constrained and values ranging from $5 \times 10^{-18} \ \mathrm{s^{-1}}$ to $5 \times 10^{-16} \ \mathrm{s^{-1}}$ are measured in dense cores \citep{Padovani2022}. A lower ionization rate would imply fewer ions and thus possibly a more efficient dust concentration, although less charges would be available for dust grains to carry.

In summary, specific conditions are needed for strong clumping that are not systemically met. However, the diversity of physical conditions (level of turbulence and magnetization, cosmic-ray ionization rate) encountered in dense core environments and protostellar envelopes make this mechanism a promising pathway to the formation of the large dust grains observed, at least in certain cases.

\subsubsection{Toward a coagulation instability}
\label{Section coag instability}
\begin{figure}
\centering
\includegraphics[width=0.5\textwidth]{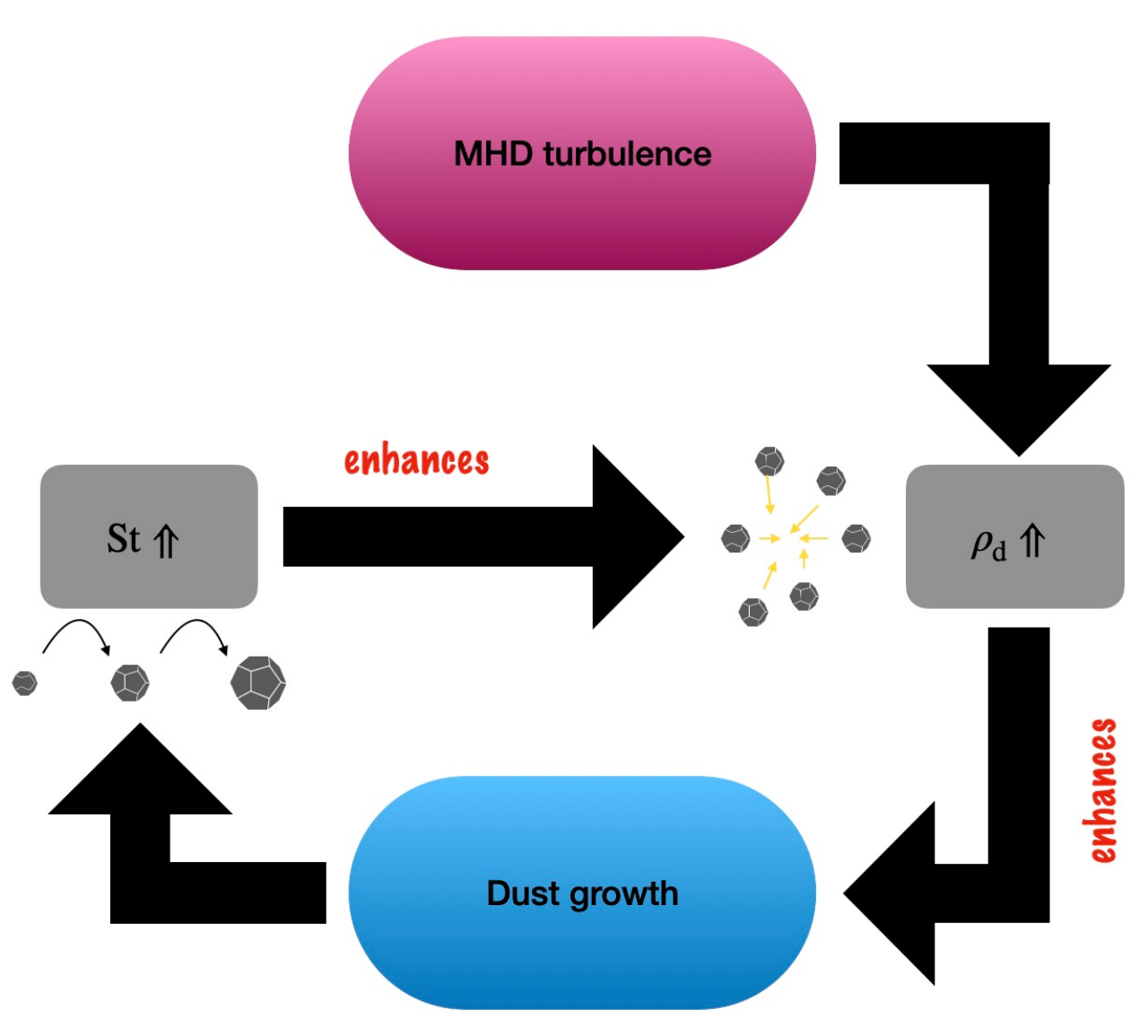}

\caption{Schematic illustration of a possible coagulation instability in turbulent magnetized dense cores. Dust concentration leads to a more rapid dust growth, which in turn results in larger grains prone to stronger concentration.}

\label{coag insta cartoon}

\end{figure}

In this section, we mention the potential development of a "coagulation instability" as defined in \cite{Tominaga2021,Tominaga2022} except that the said instability was studied in protoplanetary disks and resulted from a combination of dust growth and radial drift. In the present work, the instability would be driven by the interplay between magnetic clumping and dust growth. \newline 
On one hand, an increase in dust density as a result of magnetic clumping leads to a lower coagulation timescale. On the other hand, as dust grains grow in size (and in Stokes number), they decouple more efficiently from the gas. As showed in Sect. \ref{Section impact of grain size} and discussed in Sect. \ref{turb as a good candidate}, larger grains concentrate more strongly and reach higher density fluctuations (in the regime where the parametric instability can develop). In turn, a higher density further reduces the coagulation timescale, allowing dust grains to grow faster and clump even more efficiently, etc. The positive feedback of dust growth on dust concentration is what triggers the instability. A  very simple illustration of this concept is displayed in Fig. \ref{coag insta cartoon}. This leads to a runaway process that could allow for a very rapid growth and the formation of large grains over reasonable timescales. However, a point that remains to be elucidated is the time that would be needed to trigger this instability.

 Again, if the parametric-like instability does not overcome magnetic drag due to transverse-to-longitudinal magnetic field ratio being too low (case $\beta = 0.1$ in Fig. \ref{beta ni rhod average}), larger grains do not exhibit a higher level of clumping and the coagulation instability would thus not operate. Interestingly, local magnetic flux expulsion observed in 3D simulations of collapsing dense cores would induce inhomogeneous dust clumping and growth. Indeed, in regions of lower magnetization, the coagulation instability would be at play and dust growth would proceed very rapidly, resulting in fine in a spatial segregation/sorting of dust by size.

\subsection{Absence of convergence in the dust density}
\label{Section CV and caveats}

We show in Fig. \ref{CV test} a convergence test by plotting the dust and gas average density fluctuations with increasing resolution. While resolution has no effect on the gas density, convergence is not reached in the dust even for a number of cells as high as $\mathrm{NX} = 8192$. This results from the absence of any pressure in the dust momentum equation (Eq. \ref{final dust mom eq}), which enables the formation of smaller and sharper clumps. This further supports the capacity of magnetic effects to induce substantial dust clumping, which could be significantly higher than that presented in this work. Note that similar trends are observed in simulations of instabilities involving significant levels of dust concentration when the dust fluid is considered as pressureless \citep[for instance with the streaming instability, see][]{Johansen2007,LLambay2019}. \newline However, on sufficiently small scale, we would expect dust velocity dispersion to act as a pressure and halt further increase in dust density, even in absence of grain-grain collisions. See the work of \cite{Garaud2004,Hersant2009,Lynch&Laibe2024} for more information on the relation between velocity dispersion and pressure for a dust fluid. External forces such as gravity or magnetic forces, turbulent converging flows or even residual friction with gas (if some of it is being dragged along) could provide such velocity dispersion. The scale on which a pressure would emerge remains unclear. Marginally coupled dust grains could even concentrate to the point of gravitational instability if not opposed by pressure before, leading potentially to early planetesimal formation. \newline
In that regard, we test the robustness of our results by showing in
Fig. \ref{Fig: dust pressure} the impact of a dust pressure added a posteriori in the corresponding momentum equation. We parametrize the dust sound speed as a fraction of that of the gas. We see no effects on dust density fluctuations for values $c_\mathrm{s,d} < 0.5 c_\mathrm{s}$ for the chosen spatial resolution (NX =4096). For the highest value $c_\mathrm{s,d} =  c_\mathrm{s}$, the magnetic clumping is suppressed to a certain extent. However, it is reasonable to think that such a high incompressibility in the dust is unrealistic within our setup. To show that, we use the subgrid model of \cite{Ormel2007} which provides velocity dispersions for dust grains coupled to turbulent vortices. For a dust grain of Stokes number St, it gives:
\begin{equation}
        \Delta v_\mathrm{d} \simeq \sigma_\mathrm{v,d} = v_\mathrm{g} \sqrt{1.97 \mathrm{St}} = \sqrt{\alpha}c_\mathrm{s}\sqrt{1.97 \mathrm{St}},
\end{equation}
where $\alpha \simeq 1$ is a suitable measure of the (subsonic) turbulence intensity in dense cores. Velocity dispersions equal to $\sigma_\mathrm{v,d} \simeq c_\mathrm{s,d} =\left[0.5 c_\mathrm{s},c_\mathrm{s}\right]$ are associated with St = $\left[0.13,0.51\right]$, which are values above those explored in this work (the fiducial value being St=0.01, see Table \ref{table2}). Those simple estimates suggest that the dust concentration measured in the simulations should hold even in presence of a dust pressure in the equations. In addition, one has to keep in mind that this subgrid model ignores magnetic effects and dust backreaction. Accounting for mass loading of dust particles, we expect gas turbulence to be attenuated and thus dispersion velocities to be further reduced \citep{Tominaga2025,Carrera2025a}.

\subsection{Caveats}
A major caveat is the one dimensional nature of our simulations because 1D and 3D turbulence are known to behave differently. On one hand, geometrical effects in 3D simulations could reduce dust concentration. On the other hand, turbulent solenoidal modes would produce vortices known to be capable of trapping solid particles. Indeed particles are centrifuged out of turbulent eddies and concentrate in regions of low vorticity. This happens when the local Stokes number reaches unity \citep{Squires1991}. \newline 
As already mentioned throughout the text, there is a need for a more realistic chemical network to properly compute grain charges and ion abundances. To do so, a distribution of dust sizes will have to be implemented along with inertialess electrons in the equations, which we will do in a future work. In addition, a numerical treatment of dust growth (coagulation and fragmentation) will have to be included to follow self-consistently the formation of large grains.

\section{Conclusion}
 \label{Section conclusion}

In this work, we have investigated dust dynamics and clumping in the dense ISM ($n_\mathrm{H} = 10^4 - 10^8 \ \mathrm{cm^{-3}}$) performing multifluid 1D simulations of magnetized dust and neutral gas within two different MHD regimes:

\begin{enumerate}
    \item Dusty ideal MHD: dust grains are assumed to be perfectly coupled to magnetic field lines.
    \item Dusty non-ideal MHD: ions (whose inertia is neglected) are added in the equations along with ion-neutral friction, introducing an Ohmic dissipation and a dispersive Hall effect in the induction equation as well as an additional "magnetic drag" in the momentum equations that recouples dust and gas.
\end{enumerate}

We first propagated non-linear circularly polarized Alfvén waves and identified a mechanism similar to the parametric instability to be an efficient pathway to dust concentration. Strong dust clumping is obtained in the dusty ideal regime on scales where hydrodynamical dust-gas friction is negligible. Lower but remarkable levels of dust concentration are maintained in the dusty non-ideal regime despite non-ideal MHD effects (magnetic drag) imperilling the survival of dust clumps. Short wavelengths and sufficiently high wave amplitudes (i.e., high transverse-to-longitudinal magnetic ratio $B_\perp / B_\parallel$) are needed to achieve significant clumping, conditions that are met and sustained when considering a turbulent cascade driven on large scale. \newline
Turning then to simulations of 1D driven turbulence, we first described how dust clumps form. They develop in shocks, regions of high gas density where the dust can easily reach super-Alfvénic regime and distort the background longitudinal magnetic field lines to produce transverse components on small scales. We then investigated the influence of four different parameters on dust and gas density fluctuations, namely dust size $s_\mathrm{d}$, longitudinal $\mathcal{M}_\parallel$ and transverse $\mathcal{M}_\perp$ Mach numbers, background gas density $n_\mathrm{H}$ and plasma parameter $\beta$. Significant dust concentration, larger than that obtained with pure hydrodynamical simulations is achieved because of small scale compressive magnetic effects. There is a competition between the parametric-like instability which promotes dust concentration, and magnetic drag which hinders it. The former is sensitive to the transverse-to-longitudinal magnetic field ratio $B_\perp / B_\parallel$ while the latter can be weakened by reducing the ion abundance $n_\mathrm{i}$. As a consequence, stronger dust clumping is expected in regions of intense turbulence ($\mathcal{M}_\perp$ controls $B_\perp$) and lower (but not too low) magnetization ($\beta$ controls $B_\parallel$). Indeed, for a high magnetization, field lines are more difficult to bend and lower $B_\perp / B_\parallel$ are produced for a given level of turbulence. \newline
We found that a substantial fraction of the dust mass is easily and rapidly concentrated to a density more than 10 times its initial value, provided that $\beta \geq 0.7$ with subsonic turbulence and dust size $s_\mathrm{d} \geq 1 \ \mathrm{\mu m}$, making this novel mechanism a promising candidate for the formation of large grains in-situ in protostellar envelopes.

 \begin{acknowledgements}
We thank the referee who helped us greatly in improving the clarity and relevancy of this paper. We thank the consortium and ERC (European Research Council) synergy grant ECOGAL (grant 855130) for their financial support, and members for their contribution through insightful ideas and remarks.
 
\end{acknowledgements}

\bibliographystyle{aa}
\bibliography{ref}

\newcommand{\noop}[1]{}
\begin{thebibliography}{88}
\expandafter\ifx\csname natexlab\endcsname\relax\def\natexlab#1{#1}\fi

\bibitem[{{Akimkin} {et~al.}(2023){Akimkin}, {Ivlev}, {Caselli}, {Gong}, \& {Silsbee}}]{Akimkin2023}
{Akimkin}, V., {Ivlev}, A.~V., {Caselli}, P., {Gong}, M., \& {Silsbee}, K. 2023, \href{http://dx.doi.org/10.3847/1538-4357/ace2c5}{\color{magenta}\apj}, \href{https://ui.adsabs.harvard.edu/abs/2023ApJ...953...72A}{953, 72}

\bibitem[{{Barenblatt}(1952)}]{Barenb}
{Barenblatt}, G. 1952, Prikladnaya Matematika i Mekhanika, 16, 16

\bibitem[{{Barranco} \& {Goodman}(1998)}]{Barranco1998}
{Barranco}, J.~A. \& {Goodman}, A.~A. 1998, \href{http://dx.doi.org/10.1086/306044}{\color{magenta}\apj}, \href{https://ui.adsabs.harvard.edu/abs/1998ApJ...504..207B}{504, 207}

\bibitem[{{Ben{\'\i}tez-Llambay} {et~al.}(2019){Ben{\'\i}tez-Llambay}, {Krapp}, \& {Pessah}}]{LLambay2019}
{Ben{\'\i}tez-Llambay}, P., {Krapp}, L., \& {Pessah}, M.~E. 2019, \href{http://dx.doi.org/10.3847/1538-4365/ab0a0e}{\color{magenta}\apjs}, \href{https://ui.adsabs.harvard.edu/abs/2019ApJS..241...25B}{241, 25}

\bibitem[{{Birnstiel}(2024)}]{Birnstiel2024}
{Birnstiel}, T. 2024, \href{http://dx.doi.org/10.1146/annurev-astro-071221-052705}{\color{magenta}\araa}, \href{https://ui.adsabs.harvard.edu/abs/2024ARA&A..62..157B}{62, 157}

\bibitem[{{Cacciapuoti} {et~al.}(2025){Cacciapuoti}, {Testi}, {Maury}, {Chandler}, {Sakai}, {Ceccarelli}, {Codella}, {De Simone}, {Podio}, {Sabatini}, {Bianchi}, {Macias}, {Miotello}, {Toci}, {Loinard}, {Johnstone}, {Liu}, {Aikawa}, {Shirley}, {Svoboda}, {Sakai}, {Hirota}, {Viti}, {Lefloch}, {Oya}, {Ohashi}, {Feng}, {Fontani}, {Watanabe}, {Lopez-Sepulcre}, {Zhang}, {Vastel}, {Segura-Cox}, {Pineda}, {Isella}, {Klessen}, {Hennebelle}, {Molinari}, \& {Yamamoto}}]{Cacciapuoti2025}
{Cacciapuoti}, L., {Testi}, L., {Maury}, A.~J., {et~al.} 2025, \href{http://dx.doi.org/10.1051/0004-6361/202554645}{\color{magenta}\aap}, \href{https://ui.adsabs.harvard.edu/abs/2025A&A...700A.188C}{700, A188}

\bibitem[{{Cacciapuoti, L.} {et~al.}(2023){Cacciapuoti, L.}, {Macias, E.}, {Maury, A. J.}, {Chandler, C. J.}, {Sakai, N.}, {Tychoniec, Ł.}, {Viti, S.}, {Natta, A.}, {De Simone, M.}, {Miotello, A.}, {Codella, C.}, {Ceccarelli, C.}, {Podio, L.}, {Fedele, D.}, {Johnstone, D.}, {Shirley, Y.}, {Liu, B. J.}, {Bianchi, E.}, {Zhang, Z. E.}, {Pineda, J.}, {Loinard, L.}, {Ménard, F.}, {Lebreuilly, U.}, {Klessen, R. S.}, {Hennebelle, P.}, {Molinari, S.}, {Testi, L.}, \& {Yamamoto, S.}}]{Cacciapuoti2023}
{Cacciapuoti, L.}, {Macias, E.}, {Maury, A. J.}, {et~al.} 2023, \href{http://dx.doi.org/10.1051/0004-6361/202346204}{\color{magenta}A\&A}, 676, 676

\bibitem[{{Cadiou} {et~al.}(2019){Cadiou}, {Dubois}, \& {Pichon}}]{Cadiou19}
{Cadiou}, C., {Dubois}, Y., \& {Pichon}, C. 2019, \href{http://dx.doi.org/10.1051/0004-6361/201834496}{\color{magenta}\aap}, \href{https://ui.adsabs.harvard.edu/abs/2019A&A...621A..96C}{621, A96}

\bibitem[{{Carpine} {et~al.}(2025){Carpine}, {Ysard}, {Maury}, \& {Jones}}]{Carpine2025}
{Carpine}, M.~A., {Ysard}, N., {Maury}, A., \& {Jones}, A. 2025, \href{http://dx.doi.org/10.1051/0004-6361/202554575}{\color{magenta}\aap}, \href{https://ui.adsabs.harvard.edu/abs/2025A&A...698A.200C}{698, A200}

\bibitem[{{Carrera} {et~al.}(2025){Carrera}, {Lim}, {Eriksson}, {Lyra}, \& {Simon}}]{Carrera2025a}
{Carrera}, D., {Lim}, J., {Eriksson}, L. E.~J., {Lyra}, W., \& {Simon}, J.~B. 2025, \href{http://dx.doi.org/10.1051/0004-6361/202554100}{\color{magenta}\aap}, \href{https://ui.adsabs.harvard.edu/abs/2025A&A...696L..23C}{696, L23}

\bibitem[{{Choudhury} {et~al.}(2021){Choudhury}, {Pineda}, {Caselli}, {Offner}, {Rosolowsky}, {Friesen}, {Redaelli}, {Chac{\'o}n-Tanarro}, {Shirley}, {Punanova}, \& {Kirk}}]{Choudhury2021}
{Choudhury}, S., {Pineda}, J.~E., {Caselli}, P., {et~al.} 2021, \href{http://dx.doi.org/10.1051/0004-6361/202039897}{\color{magenta}\aap}, \href{https://ui.adsabs.harvard.edu/abs/2021A&A...648A.114C}{648, A114}

\bibitem[{{Commer{\c{c}}on} {et~al.}(2023){Commer{\c{c}}on}, {Lebreuilly}, {Price}, {Lovascio}, {Laibe}, \& {Hennebelle}}]{Commerçon2023}
{Commer{\c{c}}on}, B., {Lebreuilly}, U., {Price}, D.~J., {et~al.} 2023, \href{http://dx.doi.org/10.1051/0004-6361/202245141}{\color{magenta}\aap}, \href{https://ui.adsabs.harvard.edu/abs/2023A&A...671A.128C}{671, A128}

\bibitem[{{Crutcher}(2012)}]{Crutcher2012}
{Crutcher}, R.~M. 2012, \href{http://dx.doi.org/10.1146/annurev-astro-081811-125514}{\color{magenta}\araa}, \href{https://ui.adsabs.harvard.edu/abs/2012ARA&A..50...29C}{50, 29}

\bibitem[{{Del Zanna} {et~al.}(2001){Del Zanna}, {Velli}, \& {Londrillo}}]{DelZana2001}
{Del Zanna}, L., {Velli}, M., \& {Londrillo}, P. 2001, \href{http://dx.doi.org/10.1051/0004-6361:20000455}{\color{magenta}\aap}, \href{https://ui.adsabs.harvard.edu/abs/2001A&A...367..705D}{367, 705}

\bibitem[{{Demyk} {et~al.}(2017){Demyk}, {Meny}, {Lu}, {Papatheodorou}, {Toplis}, {Leroux}, {Depecker}, {Brubach}, {Roy}, {Nayral}, {Ojo}, {Delpech}, {Paradis}, \& {Gromov}}]{Demyk2017}
{Demyk}, K., {Meny}, C., {Lu}, X.~H., {et~al.} 2017, \href{http://dx.doi.org/10.1051/0004-6361/201629711}{\color{magenta}\aap}, \href{https://ui.adsabs.harvard.edu/abs/2017A&A...600A.123D}{600, A123}

\bibitem[{{Derby}(1978)}]{Derby1978}
{Derby}, Jr., N.~F. 1978, \href{http://dx.doi.org/10.1086/156451}{\color{magenta}\apj}, \href{https://ui.adsabs.harvard.edu/abs/1978ApJ...224.1013D}{224, 1013}

\bibitem[{{Dominik} \& {Tielens}(1997)}]{Dominik1997}
{Dominik}, C. \& {Tielens}, A.~G.~G.~M. 1997, \href{http://dx.doi.org/10.1086/303996}{\color{magenta}\apj}, \href{https://ui.adsabs.harvard.edu/abs/1997ApJ...480..647D}{480, 647}

\bibitem[{{Draine}(2004)}]{Draine2004}
{Draine}, B.~T. 2004, in The Cold Universe, \href{https://ui.adsabs.harvard.edu/abs/2004tcu..conf..213D}{213}

\bibitem[{{Draine} \& {Sutin}(1987)}]{Draine87}
{Draine}, B.~T. \& {Sutin}, B. 1987, \href{http://dx.doi.org/10.1086/165596}{\color{magenta}\apj}, \href{https://ui.adsabs.harvard.edu/abs/1987ApJ...320..803D}{320, 803}

\bibitem[{{Drazkowska}(2017)}]{Drazkowska2017}
{Drazkowska}, J. 2017, in European Planetary Science Congress, \href{https://ui.adsabs.harvard.edu/abs/2017EPSC...11..815D}{EPSC2017--815}

\bibitem[{{Elmegreen}(1987)}]{Elmegreen1987}
{Elmegreen}, B.~G. 1987, \href{http://dx.doi.org/10.1086/164907}{\color{magenta}\apj}, \href{https://ui.adsabs.harvard.edu/abs/1987ApJ...312..626E}{312, 626}

\bibitem[{{Epstein}(1924)}]{Epstein1924}
{Epstein}, P.~S. 1924, \href{http://dx.doi.org/10.1103/PhysRev.23.710}{\color{magenta}Physical Review}, \href{http://adsabs.harvard.edu/abs/1924PhRv...23..710E}{23, 710}

\bibitem[{{Galametz} {et~al.}(2019){Galametz}, {Maury}, {Valdivia}, {Testi}, {Belloche}, \& {Andr{\'e}}}]{Galametz2019}
{Galametz}, M., {Maury}, A.~J., {Valdivia}, V., {et~al.} 2019, \href{http://dx.doi.org/10.1051/0004-6361/201936342}{\color{magenta}\aap}, \href{https://ui.adsabs.harvard.edu/abs/2019A&A...632A...5G}{632, A5}

\bibitem[{{Galeev} \& {Oraevskii}(1963)}]{Galeev1963}
{Galeev}, A.~A. \& {Oraevskii}, V.~N. 1963, Soviet Physics Doklady, \href{https://ui.adsabs.harvard.edu/abs/1963SPhD....7..988G}{7, 988}

\bibitem[{{Garaud} {et~al.}(2004){Garaud}, {Barri{\`e}re-Fouchet}, \& {Lin}}]{Garaud2004}
{Garaud}, P., {Barri{\`e}re-Fouchet}, L., \& {Lin}, D.~N.~C. 2004, \href{http://dx.doi.org/10.1086/381385}{\color{magenta}\apj}, \href{https://ui.adsabs.harvard.edu/abs/2004ApJ...603..292G}{603, 292}

\bibitem[{{Goldstein}(1978)}]{Goldstein1978}
{Goldstein}, M.~L. 1978, \href{http://dx.doi.org/10.1086/155829}{\color{magenta}\apj}, \href{https://ui.adsabs.harvard.edu/abs/1978ApJ...219..700G}{219, 700}

\bibitem[{{Hendrix} \& {Keppens}(2014)}]{Hendrix2014}
{Hendrix}, T. \& {Keppens}, R. 2014, \href{http://dx.doi.org/10.1051/0004-6361/201322322}{\color{magenta}\aap}, \href{https://ui.adsabs.harvard.edu/abs/2014A&A...562A.114H}{562, A114}

\bibitem[{{Hennebelle}(2021)}]{Hennebelle2021}
{Hennebelle}, P. 2021, \href{http://dx.doi.org/10.1051/0004-6361/202141650}{\color{magenta}\aap}, \href{https://ui.adsabs.harvard.edu/abs/2021A&A...655A...3H}{655, A3}

\bibitem[{{Hennebelle} {et~al.}(2011){Hennebelle}, {Commer{\c{c}}on}, {Joos}, {Klessen}, {Krumholz}, {Tan}, \& {Teyssier}}]{Hennebelle2011}
{Hennebelle}, P., {Commer{\c{c}}on}, B., {Joos}, M., {et~al.} 2011, \href{http://dx.doi.org/10.1051/0004-6361/201016052}{\color{magenta}\aap}, \href{https://ui.adsabs.harvard.edu/abs/2011A&A...528A..72H}{528, A72}

\bibitem[{{Hennebelle} \& {Lebreuilly}(2023)}]{Hennebelle2023}
{Hennebelle}, P. \& {Lebreuilly}, U. 2023, \href{http://dx.doi.org/10.1051/0004-6361/202245120}{\color{magenta}\aap}, \href{https://ui.adsabs.harvard.edu/abs/2023A&A...674A.149H}{674, A149}

\bibitem[{{Hennebelle} \& {Passot}(2006)}]{Hennebelle&Passot2006}
{Hennebelle}, P. \& {Passot}, T. 2006, \href{http://dx.doi.org/10.1051/0004-6361:20053510}{\color{magenta}\aap}, \href{https://ui.adsabs.harvard.edu/abs/2006A&A...448.1083H}{448, 1083}

\bibitem[{{Hersant}(2009)}]{Hersant2009}
{Hersant}, F. 2009, \href{http://dx.doi.org/10.1051/0004-6361/200911865}{\color{magenta}\aap}, \href{https://ui.adsabs.harvard.edu/abs/2009A&A...502..385H}{502, 385}

\bibitem[{{Higashi} {et~al.}(2021){Higashi}, {Susa}, \& {Chiaki}}]{Higashi2021}
{Higashi}, S., {Susa}, H., \& {Chiaki}, G. 2021, \href{http://dx.doi.org/10.3847/1538-4357/ac01c7}{\color{magenta}\apj}, \href{https://ui.adsabs.harvard.edu/abs/2021ApJ...915..107H}{915, 107}

\bibitem[{{Hopkins} \& {Lee}(2016)}]{Hopkins2016}
{Hopkins}, P.~F. \& {Lee}, H. 2016, \href{http://dx.doi.org/10.1093/mnras/stv2745}{\color{magenta}\mnras}, \href{https://ui.adsabs.harvard.edu/abs/2016MNRAS.456.4174H}{456, 4174}

\bibitem[{{Hopkins} \& {Squire}(2018)}]{Hopkins2018}
{Hopkins}, P.~F. \& {Squire}, J. 2018, \href{http://dx.doi.org/10.1093/mnras/sty1604}{\color{magenta}\mnras}, \href{https://ui.adsabs.harvard.edu/abs/2018MNRAS.479.4681H}{479, 4681}

\bibitem[{{Ivlev} {et~al.}(2015){Ivlev}, {Padovani}, {Galli}, \& {Caselli}}]{Ivlev2015}
{Ivlev}, A.~V., {Padovani}, M., {Galli}, D., \& {Caselli}, P. 2015, \href{http://dx.doi.org/10.1088/0004-637X/812/2/135}{\color{magenta}\apj}, \href{https://ui.adsabs.harvard.edu/abs/2015ApJ...812..135I}{812, 135}

\bibitem[{{Johansen} \& {Youdin}(2007)}]{Johansen2007}
{Johansen}, A. \& {Youdin}, A. 2007, \href{http://dx.doi.org/10.1086/516730}{\color{magenta}\apj}, \href{https://ui.adsabs.harvard.edu/abs/2007ApJ...662..627J}{662, 627}

\bibitem[{{Kim} \& {Ryu}(2005)}]{Kim&Ryu2005}
{Kim}, J. \& {Ryu}, D. 2005, \href{http://dx.doi.org/10.1086/491600}{\color{magenta}\apjl}, \href{https://ui.adsabs.harvard.edu/abs/2005ApJ...630L..45K}{630, L45}

\bibitem[{{Krapp} \& {Ben{\'\i}tez-Llambay}(2020)}]{Krapp2020b}
{Krapp}, L. \& {Ben{\'\i}tez-Llambay}, P. 2020, \href{http://dx.doi.org/10.3847/2515-5172/abc7be}{\color{magenta}Research Notes of the American Astronomical Society}, \href{https://ui.adsabs.harvard.edu/abs/2020RNAAS...4..198K}{4, 198}

\bibitem[{Krapp {et~al.}(2019)Krapp, Ben{\'\i}tez-Llambay, Gressel, \& Pessah}]{Krapp2019}
Krapp, L., Ben{\'\i}tez-Llambay, P., Gressel, O., \& Pessah, M.~E. 2019, \href{http://dx.doi.org/10.3847/2041-8213/ab2596}{\color{magenta}\apjl}, 878, 878

\bibitem[{{Krapp} {et~al.}(2020){Krapp}, {Youdin}, {Kratter}, \& {Ben{\'\i}tez-Llambay}}]{Krapp2020a}
{Krapp}, L., {Youdin}, A.~N., {Kratter}, K.~M., \& {Ben{\'\i}tez-Llambay}, P. 2020, \href{http://dx.doi.org/10.1093/mnras/staa1854}{\color{magenta}\mnras}, \href{https://ui.adsabs.harvard.edu/abs/2020MNRAS.497.2715K}{497, 2715}

\bibitem[{{Lebreuilly} {et~al.}(2020){Lebreuilly}, {Commer{\c{c}}on}, \& {Laibe}}]{Lebreuilly2020}
{Lebreuilly}, U., {Commer{\c{c}}on}, B., \& {Laibe}, G. 2020, \href{http://dx.doi.org/10.1051/0004-6361/202038174}{\color{magenta}\aap}, \href{https://ui.adsabs.harvard.edu/abs/2020A&A...641A.112L}{641, A112}

\bibitem[{{Lebreuilly} {et~al.}(2023){Lebreuilly}, {Vallucci-Goy}, {Guillet}, {Lombart}, \& {Marchand}}]{Lebreuilly2023}
{Lebreuilly}, U., {Vallucci-Goy}, V., {Guillet}, V., {Lombart}, M., \& {Marchand}, P. 2023, \href{http://dx.doi.org/10.1093/mnras/stac3220}{\color{magenta}\mnras}, \href{https://ui.adsabs.harvard.edu/abs/2023MNRAS.518.3326L}{518, 3326}

\bibitem[{{Lee} {et~al.}(2017){Lee}, {Hopkins}, \& {Squire}}]{Lee&Hopkins2017}
{Lee}, H., {Hopkins}, P.~F., \& {Squire}, J. 2017, \href{http://dx.doi.org/10.1093/mnras/stx1097}{\color{magenta}\mnras}, \href{https://ui.adsabs.harvard.edu/abs/2017MNRAS.469.3532L}{469, 3532}

\bibitem[{{Lehmann} \& {Lin}(2023)}]{Lehmann2023}
{Lehmann}, M. \& {Lin}, M.-K. 2023, \href{http://dx.doi.org/10.1093/mnras/stad1349}{\color{magenta}\mnras}, \href{https://ui.adsabs.harvard.edu/abs/2023MNRAS.522.5892L}{522, 5892}

\bibitem[{{Li} \& {Youdin}(2021)}]{LiYoudin2021}
{Li}, R. \& {Youdin}, A.~N. 2021, \href{http://dx.doi.org/10.3847/1538-4357/ac0e9f}{\color{magenta}\apj}, \href{https://ui.adsabs.harvard.edu/abs/2021ApJ...919..107L}{919, 107}

\bibitem[{{Li} {et~al.}(2011){Li}, {Krasnopolsky}, \& {Shang}}]{Li2011}
{Li}, Z.-Y., {Krasnopolsky}, R., \& {Shang}, H. 2011, \href{http://dx.doi.org/10.1088/0004-637X/738/2/180}{\color{magenta}\apj}, \href{https://ui.adsabs.harvard.edu/abs/2011ApJ...738..180L}{738, 180}

\bibitem[{{Lynch} \& {Laibe}(2024)}]{Lynch&Laibe2024}
{Lynch}, E.~M. \& {Laibe}, G. 2024, \href{http://dx.doi.org/10.1017/jfm.2024.1088}{\color{magenta}Journal of Fluid Mechanics}, \href{https://ui.adsabs.harvard.edu/abs/2024JFM..1001A..37L}{1001, A37}

\bibitem[{{Marchand} {et~al.}(2018){Marchand}, {Commer{\c{c}}on}, \& {Chabrier}}]{Marchand2018}
{Marchand}, P., {Commer{\c{c}}on}, B., \& {Chabrier}, G. 2018, \href{http://dx.doi.org/10.1051/0004-6361/201832907}{\color{magenta}\aap}, \href{https://ui.adsabs.harvard.edu/abs/2018A&A...619A..37M}{619, A37}

\bibitem[{{Marchand} {et~al.}(2021){Marchand}, {Guillet}, {Lebreuilly}, \& {Mac Low}}]{Marchand2021}
{Marchand}, P., {Guillet}, V., {Lebreuilly}, U., \& {Mac Low}, M.~M. 2021, \href{http://dx.doi.org/10.1051/0004-6361/202040077}{\color{magenta}\aap}, \href{https://ui.adsabs.harvard.edu/abs/2021A&A...649A..50M}{649, A50}

\bibitem[{{Marchand} {et~al.}(2016){Marchand}, {Masson}, {Chabrier}, {Hennebelle}, {Commer{\c c}on}, \& {Vaytet}}]{Marchand2016}
{Marchand}, P., {Masson}, J., {Chabrier}, G., {et~al.} 2016, \href{http://dx.doi.org/10.1051/0004-6361/201526780}{\color{magenta}\aap}, \href{http://adsabs.harvard.edu/abs/2016A%26A...592A..18M}{592, A18}

\bibitem[{{Marchand} {et~al.}(2019){Marchand}, {Tomida}, {Commer{\c{c}}on}, \& {Chabrier}}]{Marchand2019}
{Marchand}, P., {Tomida}, K., {Commer{\c{c}}on}, B., \& {Chabrier}, G. 2019, \href{http://dx.doi.org/10.1051/0004-6361/201936215}{\color{magenta}\aap}, \href{https://ui.adsabs.harvard.edu/abs/2019A&A...631A..66M}{631, A66}

\bibitem[{{Mathis} {et~al.}(1977){Mathis}, {Rumpl}, \& {Nordsieck}}]{Mathis1977}
{Mathis}, J.~S., {Rumpl}, W., \& {Nordsieck}, K.~H. 1977, \href{http://dx.doi.org/10.1086/155591}{\color{magenta}\apj}, \href{http://adsabs.harvard.edu/abs/1977ApJ...217..425M}{217, 425}

\bibitem[{{Mattsson} {et~al.}(2019){Mattsson}, {Bhatnagar}, {Gent}, \& {Villarroel}}]{Mattsson2019}
{Mattsson}, L., {Bhatnagar}, A., {Gent}, F.~A., \& {Villarroel}, B. 2019, \href{http://dx.doi.org/10.1093/mnras/sty3369}{\color{magenta}\mnras}, \href{https://ui.adsabs.harvard.edu/abs/2019MNRAS.483.5623M}{483, 5623}

\bibitem[{{McKee} \& {Ostriker}(2007)}]{McKee2007}
{McKee}, C.~F. \& {Ostriker}, E.~C. 2007, \href{http://dx.doi.org/10.1146/annurev.astro.45.051806.110602}{\color{magenta}\araa}, \href{https://ui.adsabs.harvard.edu/abs/2007ARA&A..45..565M}{45, 565}

\bibitem[{{Moseley} \& {Teyssier}(2025)}]{moseley2025}
{Moseley}, E.~R. \& {Teyssier}, R. 2025, \href{http://dx.doi.org/10.1093/mnras/staf1260}{\color{magenta}\mnras}, \href{https://ui.adsabs.harvard.edu/abs/2025MNRAS.542.1011M}{542, 1011}

\bibitem[{{Moseley} {et~al.}(2023){Moseley}, {Teyssier}, \& {Draine}}]{Moseley2023}
{Moseley}, E.~R., {Teyssier}, R., \& {Draine}, B.~T. 2023, \href{http://dx.doi.org/10.1093/mnras/stac3231}{\color{magenta}\mnras}, \href{https://ui.adsabs.harvard.edu/abs/2023MNRAS.518.2825M}{518, 2825}

\bibitem[{{Nakano} {et~al.}(2002){Nakano}, {Nishi}, \& {Umebayashi}}]{Nakano2002}
{Nakano}, T., {Nishi}, R., \& {Umebayashi}, T. 2002, \href{http://dx.doi.org/10.1086/340587}{\color{magenta}\apj}, \href{https://ui.adsabs.harvard.edu/abs/2002ApJ...573..199N}{573, 199}

\bibitem[{{Nishi} {et~al.}(1991){Nishi}, {Nakano}, \& {Umebayashi}}]{Nishi1991}
{Nishi}, R., {Nakano}, T., \& {Umebayashi}, T. 1991, \href{http://dx.doi.org/10.1086/169682}{\color{magenta}\apj}, \href{https://ui.adsabs.harvard.edu/abs/1991ApJ...368..181N}{368, 181}

\bibitem[{{Ormel} \& {Cuzzi}(2007)}]{Ormel2007}
{Ormel}, C.~W. \& {Cuzzi}, J.~N. 2007, \href{http://dx.doi.org/10.1051/0004-6361:20066899}{\color{magenta}\aap}, \href{https://ui.adsabs.harvard.edu/abs/2007A&A...466..413O}{466, 413}

\bibitem[{{Ormel} {et~al.}(2009){Ormel}, {Paszun}, {Dominik}, \& {Tielens}}]{Ormel2009}
{Ormel}, C.~W., {Paszun}, D., {Dominik}, C., \& {Tielens}, A.~G.~G.~M. 2009, \href{http://dx.doi.org/10.1051/0004-6361/200811158}{\color{magenta}\aap}, \href{https://ui.adsabs.harvard.edu/abs/2009A&A...502..845O}{502, 845}

\bibitem[{{Padovani} {et~al.}(2022){Padovani}, {Bialy}, {Galli}, {Ivlev}, {Grassi}, {Scarlett}, {Rehill}, {Zammit}, {Fursa}, \& {Bray}}]{Padovani2022}
{Padovani}, M., {Bialy}, S., {Galli}, D., {et~al.} 2022, \href{http://dx.doi.org/10.1051/0004-6361/202142560}{\color{magenta}\aap}, \href{https://ui.adsabs.harvard.edu/abs/2022A&A...658A.189P}{658, A189}

\bibitem[{{Pattle} {et~al.}(2023){Pattle}, {Fissel}, {Tahani}, {Liu}, \& {Ntormousi}}]{Pattle2023}
{Pattle}, K., {Fissel}, L., {Tahani}, M., {Liu}, T., \& {Ntormousi}, E. 2023, in Astronomical Society of the Pacific Conference Series, Vol. 534, Protostars and Planets VII, ed. S.~{Inutsuka}, Y.~{Aikawa}, T.~{Muto}, K.~{Tomida}, \& M.~{Tamura}, \href{https://ui.adsabs.harvard.edu/abs/2023ASPC..534..193P}{193}

\bibitem[{{Pinto} \& {Galli}(2008)}]{Pinto2008}
{Pinto}, C. \& {Galli}, D. 2008, \href{http://dx.doi.org/10.1051/0004-6361:20078819}{\color{magenta}\aap}, \href{https://ui.adsabs.harvard.edu/abs/2008A&A...484...17P}{484, 17}

\bibitem[{{Price} \& {Federrath}(2010)}]{Price&Federrath2010}
{Price}, D.~J. \& {Federrath}, C. 2010, \href{http://dx.doi.org/10.1111/j.1365-2966.2010.16810.x}{\color{magenta}\mnras}, \href{https://ui.adsabs.harvard.edu/abs/2010MNRAS.406.1659P}{406, 1659}

\bibitem[{{Sadavoy} {et~al.}(2019){Sadavoy}, {Stephens}, {Myers}, {Looney}, {Tobin}, {Kwon}, {Commer{\c{c}}on}, {Segura-Cox}, {Henning}, \& {Hennebelle}}]{Sadavoy2019}
{Sadavoy}, S.~I., {Stephens}, I.~W., {Myers}, P.~C., {et~al.} 2019, \href{http://dx.doi.org/10.3847/1538-4365/ab4257}{\color{magenta}\apjs}, \href{https://ui.adsabs.harvard.edu/abs/2019ApJS..245....2S}{245, 2}

\bibitem[{{Shu} {et~al.}(1987){Shu}, {Adams}, \& {Lizano}}]{Shu1987}
{Shu}, F.~H., {Adams}, F.~C., \& {Lizano}, S. 1987, \href{http://dx.doi.org/10.1146/annurev.aa.25.090187.000323}{\color{magenta}\araa}, \href{https://ui.adsabs.harvard.edu/abs/1987ARA&A..25...23S}{25, 23}

\bibitem[{{Silsbee} {et~al.}(2022){Silsbee}, {Akimkin}, {Ivlev}, {Testi}, {Gong}, \& {Caselli}}]{Silsbee2022}
{Silsbee}, K., {Akimkin}, V., {Ivlev}, A.~V., {et~al.} 2022, \href{http://dx.doi.org/10.3847/1538-4357/ac978b}{\color{magenta}\apj}, \href{https://ui.adsabs.harvard.edu/abs/2022ApJ...940..188S}{940, 188}

\bibitem[{{Squire} \& {Hopkins}(2018)}]{Squire2018}
{Squire}, J. \& {Hopkins}, P.~F. 2018, \href{http://dx.doi.org/10.1093/mnras/sty854}{\color{magenta}\mnras}, \href{https://ui.adsabs.harvard.edu/abs/2018MNRAS.477.5011S}{477, 5011}

\bibitem[{{Squires} \& {Eaton}(1991)}]{Squires1991}
{Squires}, K.~D. \& {Eaton}, J.~K. 1991, \href{http://dx.doi.org/10.1017/S0022112091002276}{\color{magenta}Journal of Fluid Mechanics}, \href{https://ui.adsabs.harvard.edu/abs/1991JFM...226....1S}{226, 1}

\bibitem[{{Testi} {et~al.}(2014){Testi}, {Birnstiel}, {Ricci}, {Andrews}, {Blum}, {Carpenter}, {Dominik}, {Isella}, {Natta}, {Williams}, \& {Wilner}}]{Testi2014}
{Testi}, L., {Birnstiel}, T., {Ricci}, L., {et~al.} 2014, in Protostars and Planets VI, ed. H.~{Beuther}, R.~S. {Klessen}, C.~P. {Dullemond}, \& T.~{Henning}, \href{https://ui.adsabs.harvard.edu/abs/2014prpl.conf..339T}{339}

\bibitem[{{Tominaga} {et~al.}(2021){Tominaga}, {Inutsuka}, \& {Kobayashi}}]{Tominaga2021}
{Tominaga}, R.~T., {Inutsuka}, S.-i., \& {Kobayashi}, H. 2021, \href{http://dx.doi.org/10.3847/1538-4357/ac173a}{\color{magenta}\apj}, \href{https://ui.adsabs.harvard.edu/abs/2021ApJ...923...34T}{923, 34}

\bibitem[{{Tominaga} \& {Tanaka}(2025)}]{Tominaga2025}
{Tominaga}, R.~T. \& {Tanaka}, H. 2025, \href{http://dx.doi.org/10.3847/1538-4357/adbbca}{\color{magenta}Astrophysical Journal}, \href{https://ui.adsabs.harvard.edu/abs/2025ApJ...983...15T}{983, 15}

\bibitem[{{Tominaga} {et~al.}(2022){Tominaga}, {Tanaka}, {Kobayashi}, \& {Inutsuka}}]{Tominaga2022}
{Tominaga}, R.~T., {Tanaka}, H., {Kobayashi}, H., \& {Inutsuka}, S.-i. 2022, \href{http://dx.doi.org/10.3847/1538-4357/ac97e8}{\color{magenta}\apj}, \href{https://ui.adsabs.harvard.edu/abs/2022ApJ...940..152T}{940, 152}

\bibitem[{{Tricco} {et~al.}(2017){Tricco}, {Price}, \& {Laibe}}]{Tricco2017}
{Tricco}, T.~S., {Price}, D.~J., \& {Laibe}, G. 2017, \href{http://dx.doi.org/10.1093/mnrasl/slx096}{\color{magenta}\mnras}, \href{https://ui.adsabs.harvard.edu/abs/2017MNRAS.471L..52T}{471, L52}

\bibitem[{{Tsukamoto} {et~al.}(2021){Tsukamoto}, {Machida}, \& {Inutsuka}}]{Tsukamoto2021}
{Tsukamoto}, Y., {Machida}, M.~N., \& {Inutsuka}, S.-i. 2021, \href{http://dx.doi.org/10.3847/2041-8213/ac2b2f}{\color{magenta}\apjl}, \href{https://ui.adsabs.harvard.edu/abs/2021ApJ...920L..35T}{920, L35}

\bibitem[{{Tsukamoto} \& {Okuzumi}(2022)}]{Tuskamoto2022}
{Tsukamoto}, Y. \& {Okuzumi}, S. 2022, \href{http://dx.doi.org/10.3847/1538-4357/ac7b7b}{\color{magenta}\apj}, \href{https://ui.adsabs.harvard.edu/abs/2022ApJ...934...88T}{934, 88}

\bibitem[{{Tu} {et~al.}(2022){Tu}, {Li}, \& {Lam}}]{Tu2022}
{Tu}, Y., {Li}, Z.-Y., \& {Lam}, K.~H. 2022, \href{http://dx.doi.org/10.1093/mnras/stac2030}{\color{magenta}\mnras} \href{https://ui.adsabs.harvard.edu/abs/2022MNRAS.tmp.1903T}{[\eprint[arXiv]{2207.14151}]}

\bibitem[{{Valdivia} {et~al.}(2019){Valdivia}, {Maury}, {Brauer}, {Hennebelle}, {Galametz}, {Guillet}, \& {Reissl}}]{Valdivia2019}
{Valdivia}, V., {Maury}, A., {Brauer}, R., {et~al.} 2019, \href{http://dx.doi.org/10.1093/mnras/stz2056}{\color{magenta}\mnras}, \href{https://ui.adsabs.harvard.edu/abs/2019MNRAS.488.4897V}{488, 4897}

\bibitem[{{Vallucci-Goy} {et~al.}(2024){Vallucci-Goy}, {Lebreuilly}, \& {Hennebelle}}]{Vallucci2024}
{Vallucci-Goy}, V., {Lebreuilly}, U., \& {Hennebelle}, P. 2024, \href{http://dx.doi.org/10.1051/0004-6361/202348268}{\color{magenta}\aap}, \href{https://ui.adsabs.harvard.edu/abs/2024A&A...690A..23V}{690, A23}

\bibitem[{{Waitukaitis} {et~al.}(2014){Waitukaitis}, {Lee}, {Pierson}, {Forman}, \& {Jaeger}}]{Waitukaitis2014}
{Waitukaitis}, S.~R., {Lee}, V., {Pierson}, J.~M., {Forman}, S.~L., \& {Jaeger}, H.~M. 2014, \href{http://dx.doi.org/10.1103/PhysRevLett.112.218001}{\color{magenta}\prl}, \href{https://ui.adsabs.harvard.edu/abs/2014PhRvL.112u8001W}{112, 218001}

\bibitem[{{Wang} {et~al.}(2024){Wang}, {Wang}, {Xu}, {Sanhueza}, {Liu}, {Zhang}, {Lu}, {Fontani}, {Caselli}, {Busquet}, {Tan}, {Li}, {Jackson}, {Pillai}, {Ho}, {Guzm{\'a}n}, \& {Yue}}]{Wang2024}
{Wang}, C., {Wang}, K., {Xu}, F.-W., {et~al.} 2024, \href{http://dx.doi.org/10.1051/0004-6361/202347024}{\color{magenta}\aap}, \href{https://ui.adsabs.harvard.edu/abs/2024A&A...681A..51W}{681, A51}

\bibitem[{{Whitworth} {et~al.}(2025){Whitworth}, {Srinivasan}, {Pudritz}, {Mac Low}, {Eadie}, {Palau}, {Soler}, {Smith}, {Pattle}, {Robinson}, {Pillsworth}, {Wadsley}, {Brucy}, {Lebreuilly}, {Hennebelle}, {Girichidis}, {Gent}, {Marin}, {S{\'a}nchez Valido}, {Camacho}, {Klessen}, \& {V{\'a}zquez-Semadeni}}]{Whitworth2025}
{Whitworth}, D.~J., {Srinivasan}, S., {Pudritz}, R.~E., {et~al.} 2025, \href{http://dx.doi.org/10.1093/mnras/staf901}{\color{magenta}\mnras}, \href{https://ui.adsabs.harvard.edu/abs/2025MNRAS.540.2762W}{540, 2762}

\bibitem[{{Wong} {et~al.}(2016){Wong}, {Hirashita}, \& {Li}}]{Wong2016}
{Wong}, Y. H.~V., {Hirashita}, H., \& {Li}, Z.-Y. 2016, \href{http://dx.doi.org/10.1093/pasj/psw066}{\color{magenta}\pasj}, \href{https://ui.adsabs.harvard.edu/abs/2016PASJ...68...67W}{68, 67}

\bibitem[{{Xu} \& {Bai}(2022)}]{Xu&Bai2022}
{Xu}, Z. \& {Bai}, X.-N. 2022, \href{http://dx.doi.org/10.3847/1538-4357/ac31a7}{\color{magenta}\apj}, \href{https://ui.adsabs.harvard.edu/abs/2022ApJ...924....3X}{924, 3}

\bibitem[{{Youdin} \& {Goodman}(2005)}]{Youdin2005}
{Youdin}, A.~N. \& {Goodman}, J. 2005, \href{http://dx.doi.org/10.1086/426895}{\color{magenta}\apj}, \href{http://adsabs.harvard.edu/abs/2005ApJ...620..459Y}{620, 459}

\bibitem[{{Ysard} {et~al.}(2019){Ysard}, {Koehler}, {Jimenez-Serra}, {Jones}, \& {Verstraete}}]{Ysard2019}
{Ysard}, N., {Koehler}, M., {Jimenez-Serra}, I., {Jones}, A.~P., \& {Verstraete}, L. 2019, \href{http://dx.doi.org/10.1051/0004-6361/201936089}{\color{magenta}\aap}, \href{https://ui.adsabs.harvard.edu/abs/2019A&A...631A..88Y}{631, A88}

\bibitem[{{Zhao} {et~al.}(2016){Zhao}, {Caselli}, {Li}, {Krasnopolsky}, {Shang}, \& {Nakamura}}]{Zhao2016}
{Zhao}, B., {Caselli}, P., {Li}, Z.-Y., {et~al.} 2016, \href{http://dx.doi.org/10.1093/mnras/stw1124}{\color{magenta}\mnras}, \href{http://adsabs.harvard.edu/abs/2016MNRAS.460.2050Z}{460, 2050}

\end{thebibliography}

\appendix

\section{Derivation of generalized Ohm's law}
\label{Generalized Ohm's law}

We give here the few steps leading to the expression of the electric field in the dusty non-ideal MHD regime. Starting from the ion momentum equation (\ref{ion mom eq}), one gets:
\begin{equation}
\label{elec field 1}
    \vec{E} = - \frac{\vec{v_i}}{c} \times \vec{B} - \frac{\rho_\mathrm{i} \rho \gamma_\mathrm{i,g}}{en_\mathrm{i}} \left(\vec{v} - \vec{v_\mathrm{i}}\right) = - \frac{\vec{v_i}}{c} \times \vec{B} - \frac{\|\vec{B}\|}{c\Gamma_\mathrm{i}}\left(\vec{v} - \vec{v_\mathrm{i}}\right).
\end{equation}
The electroneutrality condition for ions and dust is:
\begin{equation}
\label{electroneutrality 2}
n_\mathrm{d} Z_\mathrm{d} + n_\mathrm{i} = 0.
\end{equation}
From the total electric current $\vec{J}_\mathrm{tot} = en_\mathrm{i}\vec{v_i} + en_\mathrm{d}Z_\mathrm{d}\vec{v_\mathrm{d}}$, using Maxwell-Ampère's law (Eq. \ref{Maxwell-Ampère}) along with Eq. (\ref{electroneutrality 2}), the ion velocity ensues:
\begin{equation}
\label{ion velocity}
    \vec{v_i} = \frac{c}{e4\pi n_\mathrm{i}} \nabla \times \vec{B} + \vec{v_\mathrm{d}},
\end{equation}
which we inject into Eq. (\ref{elec field 1})

\begin{equation}
\label{elec field final}
    \vec{E} = - \frac{\vec{v_\mathrm{d}}}{c} \times \vec{B} - \frac{1}{e4\pi n_\mathrm{i}} \left(\nabla \times \vec{B} \right) \times \vec{B} + \frac{\|\vec{B}\|}{\Gamma_\mathrm{i} e4\pi n_\mathrm{i}} \left(\nabla \times \vec{B} \right) + \frac{\|\vec{B}\|}{c\Gamma_\mathrm{i}} \left(\vec{v_\mathrm{d}} - \vec{v}\right).
\end{equation}
In addition, starting from $F_\mathrm{Lor,d} = n_\mathrm{d} Z_\mathrm{d} e \left(\vec{E} + \frac{\vec{v_\mathrm{d}}}{c} \times \vec{B} \right)$ and using Eq. (\ref{elec field 1}), we can rewrite the dust Lorentz force the following way:

\begin{equation}
\begin{split}
\vec{F}_{\mathrm{Lor,d}} & = -en_\mathrm{d}Z_\mathrm{d}\frac{\vec{v}_\mathrm{i}}{c} \times \vec{B} 
- en_\mathrm{d}Z_\mathrm{d} \frac{\|\vec{B}\|}{c\Gamma_\mathrm{i}} \left(\vec{v}-\vec{v}_\mathrm{i} \right) 
+ \frac{\vec{J}_\mathrm{d}}{c} \times \vec{B} \\
& = \frac{\vec{J}_\mathrm{tot}}{c} \times \vec{B} - \frac{\vec{J}_\mathrm{d}}{c} \times \vec{B} + \frac{\vec{J}_\mathrm{d}}{c} \times \vec{B} - en_\mathrm{d}Z_\mathrm{d} \frac{\|\vec{B}\|}{c\Gamma_\mathrm{i}} \left(\vec{v}-\frac{\vec{J}_\mathrm{tot}}{en_\mathrm{i}} - \vec{v}_\mathrm{d} \right) \\
& = \frac{\vec{J}_\mathrm{tot}}{c} \times \vec{B} - \frac{\|\vec{B}\|}{c\Gamma_\mathrm{i}} \vec{J}_\mathrm{tot} - en_\mathrm{d}Z_\mathrm{d} \frac{\|\vec{B}\|}{c\Gamma_\mathrm{i}} \left(\vec{v} - \vec{v}_\mathrm{d} \right).
\end{split}
\end{equation}
Similarly for the drag felt by the gas due to ion-neutral friction, starting from Eq. (\ref{ion mom eq}) and using the ion velocity (Eq. \ref{ion velocity}):

\begin{equation}
\begin{split}
    \vec{F_\mathrm{i \rightarrow N}} &= n_\mathrm{i} e \vec{E} + \frac{\vec{J}_\mathrm{i}}{c} \times \vec{B} \\
    &= -n_\mathrm{i} e \vec{v}_\mathrm{i} \times \vec{B} - \frac{n_\mathrm{i} e \|\vec{B}\|}{c\Gamma_\mathrm{i}} \left(\vec{v} - \vec{v}_\mathrm{i} \right) + \frac{\vec{J}_\mathrm{i}}{c} \times \vec{B} \\
    &= \frac{\vec{J}_\mathrm{i}}{c} \times \vec{B} -\frac{\vec{J}_\mathrm{i}}{c} \times \vec{B} - \frac{n_\mathrm{i} e \|\vec{B}\|}{c \Gamma_\mathrm{i}} \left(\vec{v} - \frac{\vec{J}_\mathrm{tot}}{en_\mathrm{i}} - \vec{v}_\mathrm{d} \right) \\
    &= \frac{\|\vec{B}\|}{c\Gamma_\mathrm{i}} \vec{J}_\mathrm{tot} + \frac{n_\mathrm{i} e \|\vec{B}\|}{c\Gamma_\mathrm{i}} \left(\vec{v}_\mathrm{d} - \vec{v} \right) \\
    &= \frac{\|\vec{B}\|}{c\Gamma_\mathrm{i}} \vec{J}_\mathrm{tot} - \frac{e n_\mathrm{d} Z_\mathrm{d} \|\vec{B}\|}{c\Gamma_\mathrm{i}} \left(\vec{v}_\mathrm{d} - \vec{v} \right).
\end{split}
\end{equation}

\section{Dispersion relation of the dusty ideal Alfvén wave}
\label{Alfven propagation section}

\begin{figure}
\centering
    \includegraphics[width= 0.5\textwidth]{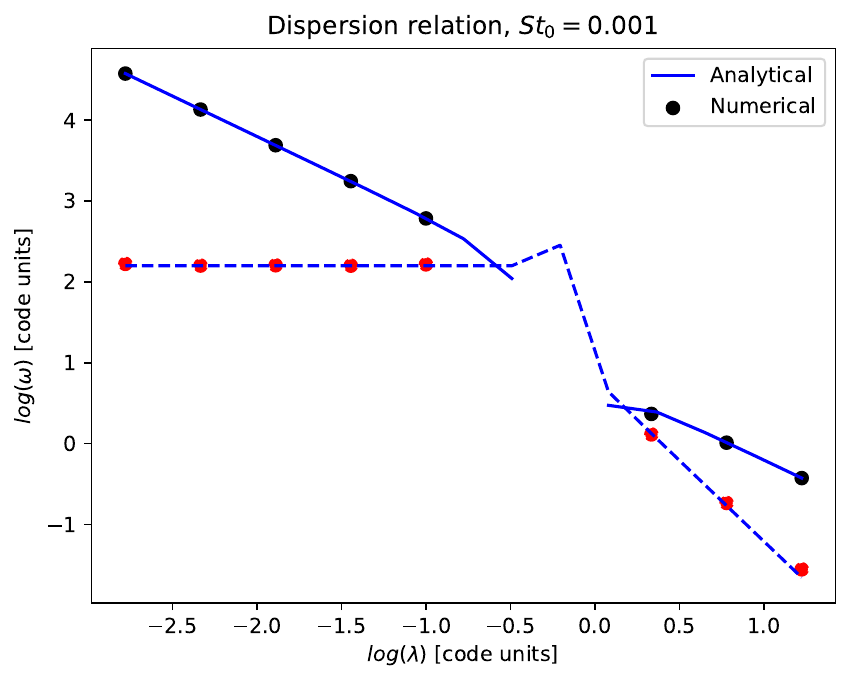}

    \caption{Dispersion relation for an Alfvén wave in a dusty magnetized medium within the dusty ideal MHD model (Sect. \ref{dusty ideal MHD}). Comparison between the analytical solution and numerical measurement from our simulation. The real part of the frequency appears as solid line and the imaginary part as dashed line. $\mathrm{St}=0.001$, $\beta=0.1$ and $\delta_B=10^{-4}$.}
\label{plot dispersion relation}
\end{figure}

We study the propagation of a linear Alfvén wave within the dusty ideal MHD model (Sect. \ref{dusty ideal MHD}) and compare in Fig. \ref{plot dispersion relation} the analytical dispersion relation with the one measured in our simulations. This analysis serves as a test for the code and as a source of physical insight.
The circularly polarized Alfvén wave is initialized as:

\begin{equation}
B_y = \delta_B B_x \sin{\left(\frac{2 \pi }{\lambda} x\right)} \\  
B_z = \delta_B B_x \cos{\left(\frac{2 \pi }{\lambda} x\right)},
\label{eq:circularly pol Alfvén wave}
\end{equation}
with the wavelength $\lambda$ being a fraction of the box length $L$. The wave amplitude is taken to be $\delta_B B_x = 10^{-4} B_x$ so that we work in the linear regime. The plasma parameter is taken to be $\beta = 0.1$ and we set the initial Stokes number to be $\mathrm{St} = 0.001$. Both gas and dust are initially moving in the transverse directions with velocity components imposed by the Alfvén mode targeted. Note that we work with code units here: we consider a uniform gas initial density $\rho_0 = 1$ and dust initial density $\rho_\mathrm{d,0} = 0.01 \rho_0$ (i.e. a dust-to-gas ratio $\epsilon = 0.01$), a uniform x-wise background magnetic field inferred from the plasma parameter $\beta$ as $B_x = c_\mathrm{s} \sqrt{4 \pi \rho_0/\beta} $, and a constant soundspeed $c_\mathrm{s} = 0.316$.

\subsection{Analytical solution}

To derive the analytical dispersion relation, we start from the gas and dust momentum equations (Eq. \ref{eq:hydro ideal}) as well as the induction equation (Eq. \ref{induction equation ideal}) and carry out a linear perturbation analysis. Since Alfvén modes are incompressible, we discard the mass conservation equations. In addition, we consider no background quantities except for $B_x$ and we work with transverse perturbations only. 
Combining the y and z components, we use the following change of variables and work with the polarization modes "+" and "-":

\begin{equation}
\label{pol mode change of var}
V_{\pm} = V_y \pm iV_z \\ 
B_{\pm} = B_y \pm iB_z.
\end{equation}
We consider the following small perturbations

\begin{align}
\label{perturbarions}
\delta v_{\pm} &=& V^{\pm} \exp \left(i \left(\omega t - kx \right) \right), \\ \nonumber
\delta v_\mathrm{d,\pm} &=& V_\mathrm{d}^{\pm} \exp \left(i \left(\omega t - kx \right) \right), \\ \nonumber
\delta b_{\pm} &=& B^{\pm} \exp \left(i \left(\omega t - kx \right) \right),
\end{align}

where $w$ and $k$ are the perturbation frequency and wavenumber. The momentum and induction equations then become:

\begin{align}
\label{perturbated eq}
i \omega V^{\pm} &=& \frac{\rho_\mathrm{d}}{\rho t_\mathrm{s}} \left(V_\mathrm{d}^{\pm} - V^{\pm} \right), \\ \nonumber
i \omega V_\mathrm{d}^{\pm} &=& -\frac{\left(V_\mathrm{d}^{\pm} - V^{\pm} \right)}{t_\mathrm{s}} - \frac{ikB_x B^{\pm}}{\rho_\mathrm{d}}, \\ \nonumber
i \omega B^{\pm} &=& -B_{x} i k V_\mathrm{d}^{\pm}.
\end{align}
It is straightforward to show that the system reduces to:

\begin{equation}
\label{matrix system} 
\left( \begin{array}{cc} i \omega+ \rho_\mathrm{d}/\rho t_\mathrm{s} & \rho_\mathrm{d} \omega / \rho t_\mathrm{s} B_x k \\
-1/t_\mathrm{s} & ikB_x/\rho_\mathrm{d} - \left(i \omega + 1/t_\mathrm{s} \right) \omega/B_x k \end{array} \right)
\left( \begin{array}{c} V^{\pm} \\ B^{\pm} \end{array} \right) = 0.
\end{equation}

Equating the matrix determinant to zero, the dispersion relation ensues:

\begin{equation}
\label{dispersion relation}
\frac{\omega^3}{k}-\frac{i\left(\rho + \rho_\mathrm{d}\right)}{k \rho t_\mathrm{s}} \omega^2 -\frac{kB_x^2}{\rho_\mathrm{d}} \omega + i \frac{kB_x^2}{\rho t_\mathrm{s}} = 0,
\end{equation}
which is a third degree polynomial. One mode is evanescent and thus does not propagate. The other two are degenerate Alfvén modes which propagate in opposite directions, which is expected since there is no Hall effect breaking the symmetry.

\subsection{Comparison with simulation}

The evolution of the frequency (real part in solid line and imaginary part in dashed line) with respect to the wavelength for the progressive Alfvén mode is plotted in Fig. \ref{plot dispersion relation} with $\mathrm{St}=0.001$, $\beta=0.1$ and $\delta_B=10^{-4}$. We compare the analytical solution with the frequencies and damping rates measured in our simulation. To ensure that the right mode is excited, we inject the associated solution of Eq. (\ref{dispersion relation}) in the set of perturbed equations Eq. (\ref{perturbated eq}) and use them to infer the initial conditions for the gas and dust velocity perturbations. 
We notice a striking match between the analytical solution and the numerical measurements. At large wavelengths, the dynamical timescale of the Alfvén wave ($t_\mathrm{a} = L/c_\mathrm{a}$) is much larger than the grain stopping time (i.e. $t_\mathrm{a} >> t_\mathrm{s}$), and we thus recover the strong coupling regime where the Alfvén velocity is determined by the gas density, $c_\mathrm{a} = \|\vec{B}\|/\sqrt{4 \pi \rho} = 1$ in code units. As we decrease the wavelength, we enter a very dissipative regime where the imaginary part becomes much higher than the real part, preventing any propagation. This is due to a resonance effect ($t_\mathrm{a} \simeq t_\mathrm{s}$) where most of the magnetic energy is lost to friction with the gas. We note that the frequency of the signal could not be measured from the corresponding simulation in this regime, and we did not display the near-zero real part from the analytical solution for the sake of clarity. Finally, at low wavelength, we see that Alfvén waves do propagate with weak dissipation, which is a feature already identified analytically in \cite{Hennebelle2023} when accounting for dust inertia. Since $t_\mathrm{a} << t_\mathrm{s}$, gas-grain collisions are scarce and Alfvén waves are carried by the grains. The Alfvén waves thus propagate at a velocity ten times higher $c_\mathrm{a,d} = \|\vec{B}\|/\sqrt{4 \pi \rho_\mathrm{d}} = 10$ as we consider a dust-to-gas ratio $\epsilon = 0.01$. 

\section{Growth rate of the parametric instability}
\label{Appendix: growth rate parametric}

\begin{figure}
\centering \includegraphics[width=0.5\textwidth]{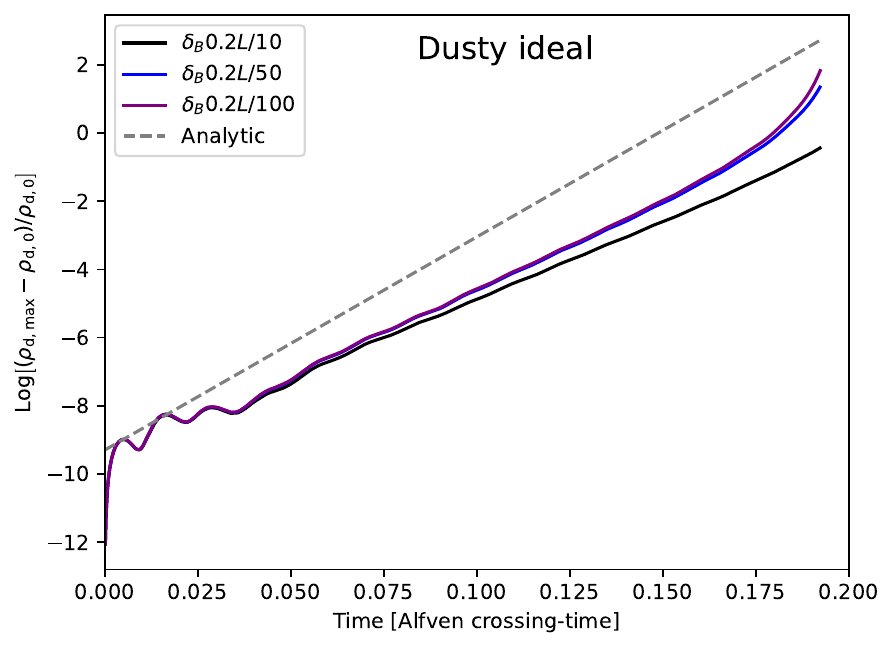}
    \caption{Comparison between the analytic growth rate of the parametric instability (grey dashed line) and numerical results of the two-fluid parametric-like instability presented in Sect. \ref{section parametric instability} in the dusty ideal MHD model for different values of the "mother" Alfvén wave wavelength (same configuration as in Fig. \ref{param instability ideal vs non-ideal}).}
    \label{fig: growth rate parametric}
\end{figure}

We aim to support the statement that the instability observed and described in Sect. \ref{section parametric instability} is very similar to the so-called parametric instability \citep{DelZana2001} but in a two-fluid configuration. To do so, we provide the analytical growth rate of the fastest growing mode derived in the standard one-fluid configuration, and compare it to the numerical results (see Fig. \ref{fig: growth rate parametric}) in the early growth phase. \newline 
For the standard parametric instability, the dispersion relation for the daughter waves (with a generic $\beta$) is \citep{Derby1978,Goldstein1978,DelZana2001}:

\begin{align}
\label{Eq: parametric dispersion relation}
   \left(\Tilde{\omega}^2 - \beta\Tilde{k}^2\right) \left(\Tilde{\omega} - \Tilde{k}\right)^2&\left(\Tilde{\omega} + \Tilde{k} + 2\right)\left(\Tilde{\omega} + \Tilde{k} - 2\right) \\ \nonumber 
&-\delta_B^2\Tilde{k}^2 \left(\Tilde{\omega} - \Tilde{k}\right) \left(\Tilde{\omega}^3 + \Tilde{k}\Tilde{\omega}^2- 3\Tilde{\omega} + \Tilde{k}\right)  = 0,
\end{align}
where the dimensionless frequency and wavenumber are respectively $\Tilde{\omega} = \omega / \omega_0$ and $\Tilde{k} = k / k_0$. We choose the "mother" Alfvén wave frequency to be $\omega_0 = c_\mathrm{a,d}k_0$ with $c_\mathrm{a,d}  = \|\vec{B}\|/\sqrt{4 \pi \rho_\mathrm{d}}$ being the dust Alfvén speed. Indeed, we account for the fact that the magnetized fluid sensitive to the instability is the pressureless dust, and set $\beta = 0$. Although the study of a two-fluid parametric instability is beyond the scope of this work, Eq. (\ref{Eq: parametric dispersion relation}) is relevant when friction between gas and dust is negligible, i.e. when considering an Alfvén wave frequency large with respect to the collision frequency of the two fluids. This condition is met in Fig. \ref{fig: growth rate parametric} where the growth rate of the fastest growing (unstable) mode computed from the roots of Eq. (\ref{Eq: parametric dispersion relation}) is displayed as a grey dashed line. \newline The analytic growth rate is little sensitive to the wavelength of the "mother" Alfvén wave and we write the dust density perturbation as $\delta \rho_\mathrm{d} \propto \exp\left(t/\tau\right)$ where we computed the growth rate $\tau \simeq 0.015$ Alfvén crossing-time. We see that there is a good match between the analytic prediction and the numerical results, reinforcing the idea that a parametric-like instability is at play. The modest deviation comes from the finite friction between the charged dust and the neutral gas and the fact that initial perturbations from the background state are induced by numerical noise (meaning that several modes are probably triggered).

\section{Numerical treatment of non-ideal MHD terms.}
\label{Diffusion test section}

\begin{figure}
\centering

    \includegraphics[width=0.5\textwidth]{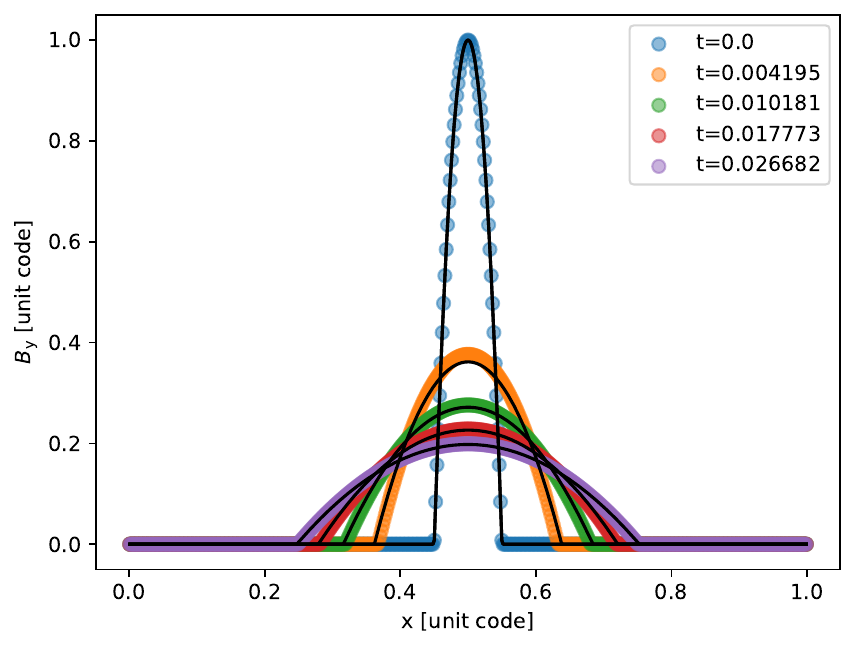}
    \caption{Comparison between analytical solution and numerical resolution of the non-linear diffusion equation given by Eq. (\ref{simplified diffusion eq}). The analytical solution \citep{Barenb} is given by Eq. (\ref{barenblatt}). Each curve shows $B_y$ profile at a given time.}
\label{Diffusion test}
\end{figure}
 
We show here the treatment made for the diffusive ohmic dissipation and the dispersive Hall effect appearing in the right-hand side of the induction equation (Eq. \ref{induction equation}). In our cartesian coordinate system, 
The Ohm-like term is written:
\begin{equation}
\label{solver ohm} 
\left( \begin{array}{c}  \partial B_x / \partial t \\  \partial B_y / \partial t \\  \partial B_z /\partial t \end{array} \right) = \left( \begin{array}{c}  0 \\ \partial_x \left(\frac{c\|\vec{B}\|}{4\pi \Gamma_\mathrm{i} en_\mathrm{i}} \partial_x B_y \right) \\ \partial_x \left(\frac{c\|\vec{B}\|}{4\pi \Gamma_\mathrm{i} en_\mathrm{i}} \partial_x B_z \right) \end{array} \right),
\end{equation}
where $\|\vec{B} \| = \sqrt{B_x^2 + B_y^2 + B_z^2}$ denotes the norm of the magnetic field.
This is a diffusion equation which is non-linear owing to the diffusion coefficients (magnetic resistivity) being space-dependent. Figure \ref{Diffusion test} provides a comparison of the numerical results with the analytical solution at different times for the following simple non-linear diffusion equation:

\begin{equation}
    \partial B_y / \partial t = \partial_x \left(2B_y \partial_x B_y \right).
\label{simplified diffusion eq}
\end{equation}
A self similar solution to this equation is the Barenblatt-Pattle solution \citep{Barenb} that writes:

\begin{equation}
    B_y\left(t,x,C \right) = (2t)^{-1/3} \left(C - \frac{1}{6} \frac{x^2}{(2t)^{2/3}} \right),
\label{barenblatt}
\end{equation}
with the constant $C = \left(B_{y,0} x_c / \sqrt{6} \right)^{2/3}$. At initial time $t_0 = C^3/2B_{y,0}^3$, we consider the following initial profile:

\begin{equation}
    B_y\left(t_0,x \right) = B_{y,0} \left(1 - \left(\frac{x}{x_c}\right)^2 \right).
\end{equation}
There is a great agreement between the numerical evolution of the profile and the analytical prediction, showing that the solver works properly and is suited to our numerical investigation. \newline
Now, we turn to the Hall-like term which is expressed as:
\begin{equation}
\label{solver Hall} 
\left( \begin{array}{c}  \partial B_x / \partial t \\  \partial B_y / \partial t \\  \partial B_z /\partial t \end{array} \right) = \left( \begin{array}{c}  0 \\ \partial_x \left(\frac{cB_x}{4\pi en_\mathrm{i}} \partial_x B_z \right) \\ -\partial_x \left(\frac{cB_x}{4\pi en_\mathrm{i}} \partial_x B_y \right) \end{array} \right).
\end{equation}
This term cannot be treated as a diffusion equation since $B_y$ and $B_z$ are coupled. Instead, we adopt the approach presented in \cite{Marchand2018} and \cite{Marchand2019} and rewrite it in conservative form using Maxwell-Ampère equation (Eq. \ref{Maxwell-Ampère}):
\begin{equation}
\label{solver Hall conservative form} 
\left( \begin{array}{c}  \partial B_x / \partial t \\  \partial B_y / \partial t \\  \partial B_z /\partial t \end{array} \right) = \left( \begin{array}{c}  0 \\ - \partial_x \left(\frac{B_x}{en_\mathrm{i}}  J_y \right) \\ -\partial_x \left(\frac{B_x}{en_\mathrm{i}}  J_z \right) \end{array} \right),
\end{equation}
and include it as a flux in a separate Riemann solver dedicated to the induction equation (Eq. \ref{induction equation}). The Hall effect is known to introduce new dispersive waves in the system of hyperbolic equations, namely Whistler waves. The dispersive nature of this term compels us to cut-off the velocity of Whistler waves on a scale that is determined by the resolution, i.e. the cell size $\Delta x = \frac{NX}{L}$. The Whistler velocity is thus defined as:

\begin{equation}
    c_w = \frac{\eta_\mathrm{H} \pi}{2 \Delta x} + \sqrt{\left(\frac{\eta_\mathrm{H} \pi}{2 \Delta x}\right)^2 + c_a^2}
\end{equation}
where the Hall resistivity $\eta_\mathrm{H} = \frac{cB_x}{4\pi en_\mathrm{i}}$. We incorporate it in the wave fan of the Riemann solver of the magnetic field, but do not use it to update non-magnetic flow variables. As pointed out in \cite{Marchand2019}, doing otherwise would introduce significant truncation errors and would thus lead to a much lower effective resolution.

\section{Convergence test}
\label{CV test section}

\begin{figure}
\centering \includegraphics[width=0.5\textwidth]{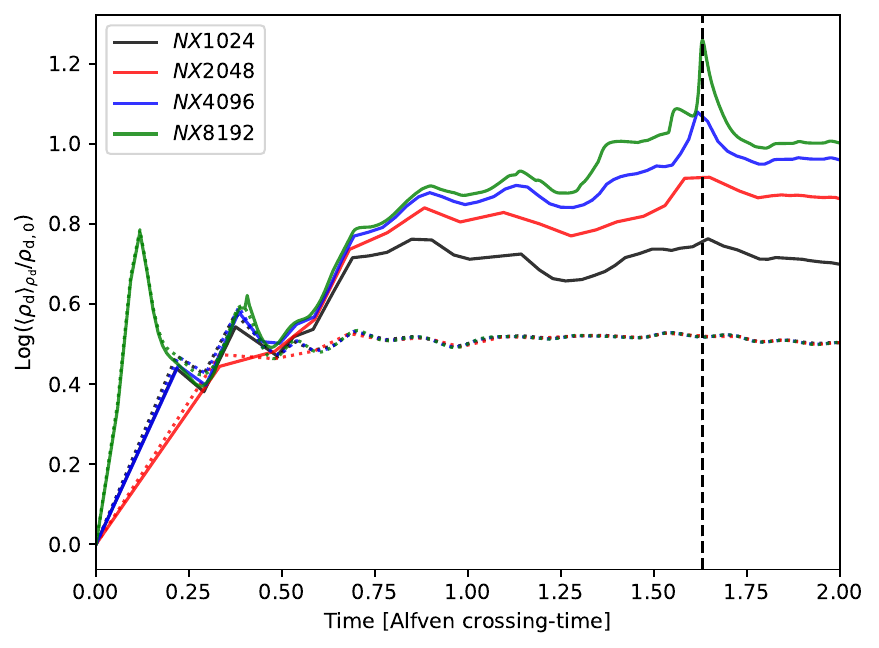}
    \caption{Convergence test. Average dust density (solid) and gas density (dotted) (as defined in Eq. \ref{averaging formulae}) as a function of time for different spatial resolutions. $\beta = 1$ and $s_\mathrm{d} = 100 \ \mathrm{\mu m}$ ($\mathrm{St = 0.1}$) were used. The vertical dashed black line indicates the time at which Fig. \ref{CV test rho field} was made. While convergence is reached for the gas, it is not for the dust.}
    \label{CV test}
\end{figure}

\begin{figure}
\centering \includegraphics[width=0.5\textwidth]{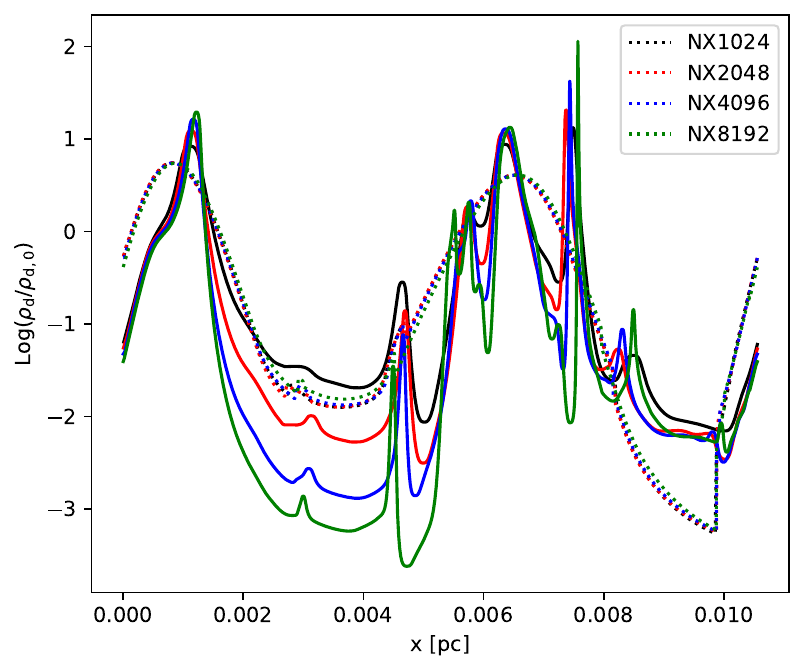}
    \caption{Dust (solid lines) and gas (dotted lines) density fluctuation field throughout the box at $t = 1.6$ (see Fig. \ref{CV test}) for different resolutions.}
    \label{CV test rho field}
\end{figure}

In Fig. \ref{CV test}, we show a convergence test by plotting the dust and gas average density fluctuations with increasing resolution. While resolution has no effect on the gas density, convergence is not reached in the dust even for a number of cells as high as $NX=8192$. This results from the absence of any pressure in the dust, which is allowed to form smaller and smaller clumps. This further supports the capacity of magnetic effects to induce substantial dust clumping, which could be significantly higher than that shown in the present work. However, on sufficiently small scale, we would expect a remnant dust velocity dispersion to act as a pressure and halt further increase in dust density.

\section{Ideal vs non-ideal dusty MHD}
\label{Ideal vs non-ideal}

\begin{figure}
\centering

    \includegraphics[width=0.5\textwidth]{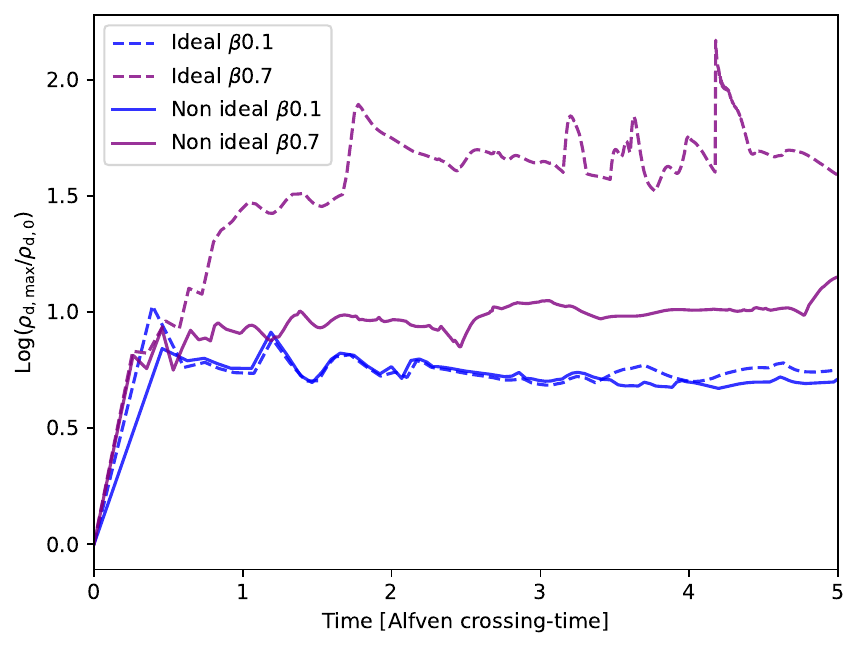}
    \caption{Comparison of time evolution of the maximum dust density within dusty ideal MHD setup (dashed line. See Sect. \ref{dusty ideal MHD}) and dusty non-ideal MHD (solid line. See Sect. \ref{Dusty non-ideal MHD}). Plasma parameter $\beta=0.1$ (blue) and $\beta=0.7$ (purple). Dust size: $s_\mathrm{d} = 10 \ \mathrm{\mu m}$.}
    \label{idealvsnonideal plot}
\end{figure}

A comparison between dusty ideal MHD model (Sect. \ref{dusty ideal MHD}) and dusty non-ideal MHD model (Sect. \ref{Dusty non-ideal MHD}) is depicted in Fig. \ref{Ideal vs non-ideal}. When plasma parameter $\beta = 0.1$, there is no difference between both models. Indeed, the fiducial transverse Mach number $\mathcal{M}_\perp = 0.99$ produces too weak transverse-to-longitudinal magnetic field ratio $B_\perp / B_\parallel$, keeping the parametric instability from inducing a substantial dust concentration beyond that of pure hydrodynamical turbulent processes. However, when going to $\beta = 0.7$, significant deviation is observed between the two models: maximum dust density fluctuations are one order of magnitude apart. Magnetic drag in Eq. (\ref{final dust mom eq}) due to the presence of ions is to be held responsible for the reduction in dust concentration within the dusty non-ideal regime. Consequently, an accurate treatment of non-ideal MHD effects cannot be ignored and have to be included in numerical simulations.

\section{How the plasma parameter $\beta$ controls the transverse-to-longitudinal magnetic ratio $B_\perp / B_\parallel$}
\label{B_ratio vs beta}

\begin{figure}
\centering

    \includegraphics[width=0.5\textwidth]{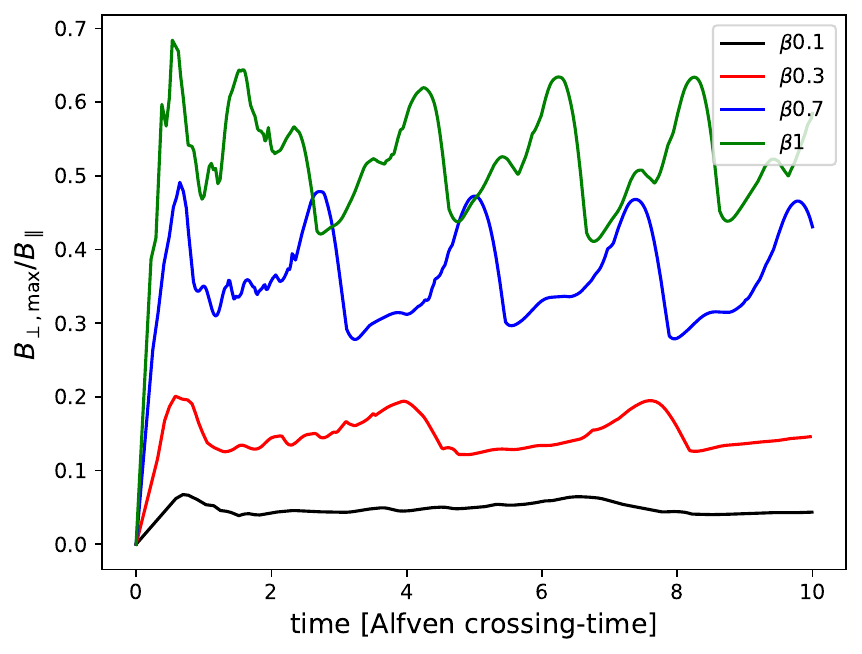}
    \caption{Maximum transverse-to-longitudinal magnetic ratio as a function of time for different values of plasma parameter $\beta$ in simulations with turbulence.}
    \label{B_perp over B_par vs beta}
\end{figure}

Figure \ref{B_perp over B_par vs beta} depicts the influence of the of the plasma parameter $\beta$ on the maximum transverse-to-longitudinal magnetic ratio $B_\perp / B_\parallel$ in simulations with turbulence. We clearly see that within our setup, for a given level of turbulence ($\mathcal{M}_\perp$), a higher $\beta$ (i.e. lower longitudinal $B_x$) leads to higher values of $B_\perp / B_\parallel$ which in turn leads to higher dust density fluctuations (see Sect. \ref{section parametric instability}).

\section{Exploring the effect of a dust pressure on density fluctuations.}
\label{Appendix: dust pressure}

\begin{figure}
\centering

    \includegraphics[width=0.5\textwidth]{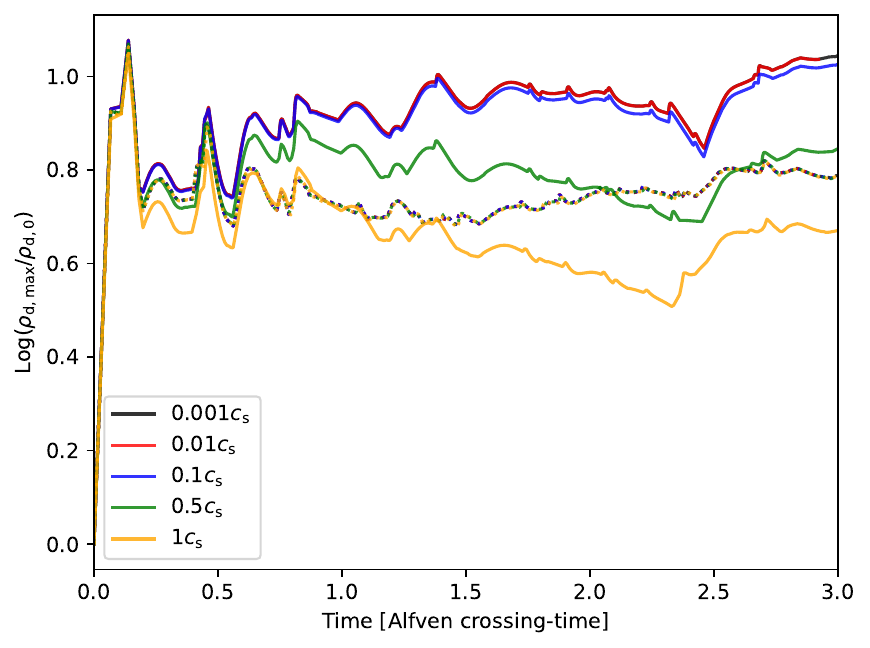}
    \caption{Maximum dust density (solid lines) and gas density (dotted lines) fluctuations as a fraction of time for different values of the dust sound speed (defined as a function of the gas sound speed $c_\mathrm{s}$) in simulations with turbulence (the red and black curves are perfectly overlain). Note that the spatial resolution is NX = 4096.}
    \label{Fig: dust pressure}
\end{figure}

Figure \ref{Fig: dust pressure} shows the impact of a dust pressure added a posteriori in the corresponding momentum equation. Parametrizing the dust sound speed as a fraction of that of the gas, we see no effects on dust density fluctuations for values $c_\mathrm{s,d} < 0.5 c_\mathrm{s}$ for the spatial resolution chosen (NX = 4096). As discussed in Sect. \ref{Section CV and caveats}, such high values seem unrealistic for the grain sizes explored in this work, meaning that our results are most likely robust.

\section{Chemical network comparison}
\label{Chemical network comparison}

\begin{figure}

\includegraphics[width=0.5\textwidth]{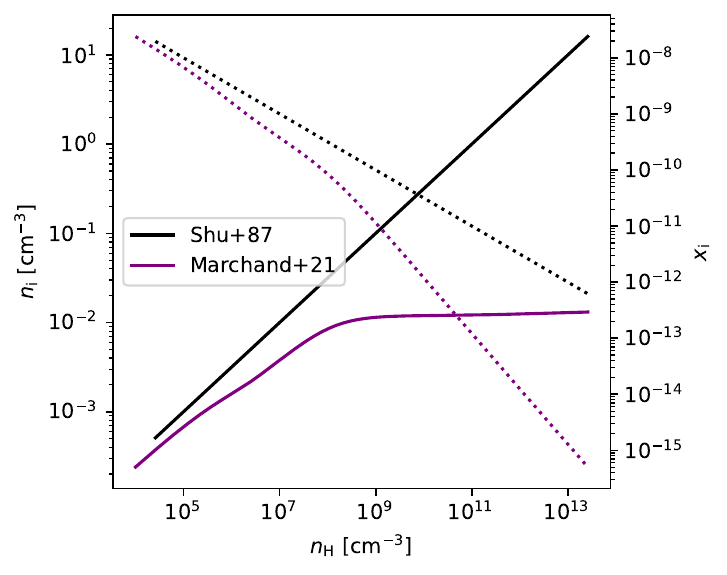}

\caption{Ion density (solid lines, left axis) and ionization fraction $n_\mathrm{i} / n_\mathrm{H}$ (dotted lines, right axis) as a function of gas density. Comparison between our chemical model \citep[Sect. \ref{chemical network},][]{Shu1987} and that of \cite{Marchand2021}. For the latter, an MRN dust size distribution was considered. For both, the ionization rate by cosmic rate is $\zeta = 10^{-17} \ \mathrm{s^{-1}}$.}
\label{Shu vs Marchand}
\end{figure}

We provide here a comparison of the chemical network used in this paper (see Sect. \ref{chemical network}) with the more sophisticated one presented in \cite{Marchand2021}. The latter includes the effects of a dust size distribution, with small grains offering a large cross section for reactions such as electron capture and ion-electron recombination at their surface. Figure \ref{Shu vs Marchand} shows the ion abundance $n_\mathrm{i}$ and ionization fraction $x_i = n_\mathrm{i} / n_\mathrm{H}$ as a function of gas density $n_\mathrm{H}$. Although both networks seem to agree at lower density ($n_\mathrm{H} < 10^5 \ \mathrm{cm^{-3}}$), discrepancies are exacerbated as density increases. 

\section{Impact of parameters on density fluctuations for a lower plasma parameter $\beta=0.1$}
\label{Dust clumping beta0.1}

\begin{figure*}

\includegraphics[width=\textwidth]{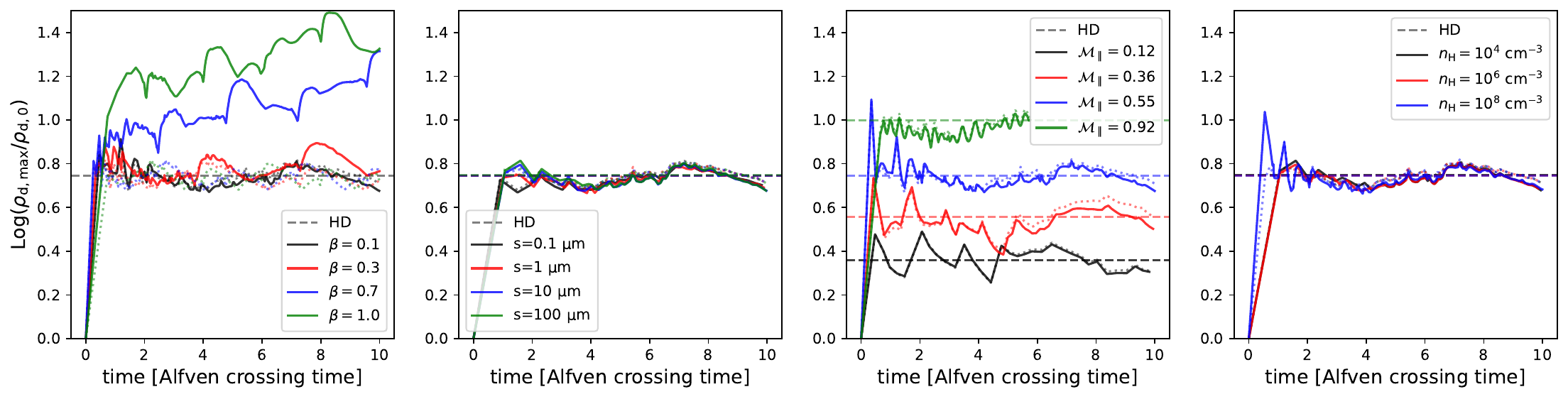}

\caption{Dust density (solid lines) and gas density (dotted lines) maximum fluctuations as a function of time for different values of the dust grain size $s_\mathrm{d}$, longitudinal Mach number (perpendicular one being varied too) $\mathcal{M}_\parallel$, gas initial density $n_\mathrm{H}$ and plasma parameter $\beta$. The dust density mean value for pure hydrodynamics simulations is displayed for reference as dashed lines. When not varied, $\beta=0.1$.}

\label{rhodvstime0.1}
\end{figure*}

\begin{figure*}

\includegraphics[width=\textwidth]{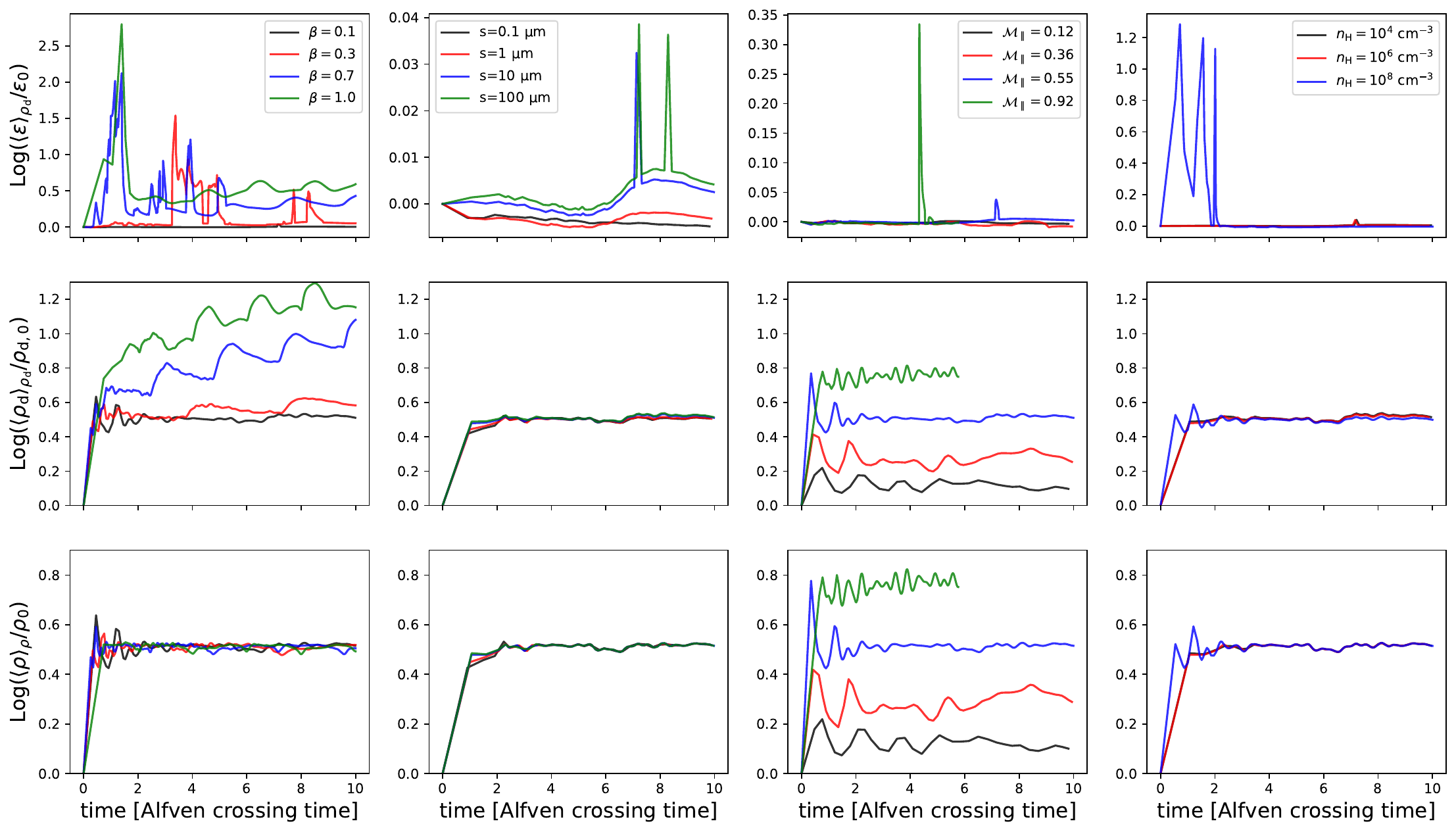}

\caption{Same as Fig. \ref{rhodvstime0.1} but depicting dust-to-gas ratio average (dust density weighted), dust density average (dust density weighted) and gas density average (gas density weighted).}

\label{average vs time 0.1}
\end{figure*}

Similarly to what was done in Sect. \ref{parametric study}, Fig. \ref{rhodvstime0.1} and Fig. \ref{average vs time 0.1} depict the influence of dust size, Mach number, background gas density and plasma parameter on maximum and average density fluctuations as a function of time, however for a lower fiducial value of plasma parameter $\beta = 0.1$. The fiducial values for the other parameters are those presented in Table \ref{table1}. As explained in Sect. \ref{section parametric instability} and Sect. \ref{turb as a good candidate}, a lower $\beta$ (i.e. a higher background $B_x$) makes the development of high transverse-to-longitudinal magnetic field ratios $B_\perp / B_\parallel$ more challenging. As a consequence, magnetic effects are incapable of pushing dust concentration levels beyond what is permitted by pure hydrodynamical effects. In addition, there is a loss of dust clumping dependency on dust size because of ion-induced magnetic drag being insensitive to this parameter, as seen in Eq. (\ref{final dust mom eq}).

\end{document}